\documentclass[%
 aip,
 amsmath,amssymb,
 reprint,%
]{revtex4-2}

\usepackage{graphicx}
\usepackage{amsmath}
\usepackage{amsfonts}
\usepackage{graphicx}
\usepackage[font=small,skip=0pt]{caption}
\usepackage{commath}
\usepackage{caption}
\usepackage{tabularx,ragged2e,booktabs,caption}
\usepackage{xcolor}
\usepackage{graphicx}
\usepackage{accents}

\makeatletter
\newcommand{\vast}{\bBigg@{4}}
\newcommand{\Vast}{\bBigg@{5}}
\makeatother

\newcommand*{\Comb}[2]{{}^{#1}C_{#2}}%
\newcommand{\ubar}[1]{\underaccent{\bar}{#1}}

\newenvironment{rcases}
  {\left.\begin{aligned}}
  {\end{aligned}\right\rbrace}
  
\newcommand*{\ditto}{---\texttt{"}---}

\begin{document}
%
%
%
%
%

\title{How zealots affect the energy cost for controlling complex social networks}

\author{Hong Chen}
\affiliation{Division of Physics and Applied Physics, School of Physical and Mathematical Sciences,
Nanyang Technological University, Singapore, 637371, Singapore}
\affiliation{Business Analytics Centre, National University of Singapore, Singapore 119613, Singapore}

\author{Ee Hou Yong \footnote{Author to whom any correspondence should be addressed.}}
\email{eehou@ntu.edu.sg}
\affiliation{Division of Physics and Applied Physics, School of Physical and Mathematical Sciences,
Nanyang Technological University, Singapore, 637371, Singapore}


\begin{abstract}
The controllability of complex networks may be applicable for understanding how to control a complex social network, where members share their opinions and influence one another. Previous works in this area have focused on controllability, energy cost, or optimization under the assumption that all nodes are compliant, passing on information neutrally without any preferences. However, the assumption on nodal neutrality should be reassessed, given that in networked social systems, some people may hold fast to their personal beliefs. By introducing some stubborn agents, or zealots, who hold steadfast to their beliefs and seek to influence others, the control energy is computed and compared against those without zealots. It was found that the presence of zealots alters the energy cost at a quadratic rate with respect to their own fixed beliefs. However, whether or not the zealots' presence increases or decreases the energy cost is affected by the interplay between different parameters such as the zealots' beliefs, number of drivers, final control time regimes, network effects, network dynamics, number and configurations of neutral nodes influenced by the zealots. For example, when a network dynamics is linear but does not have conformity behavior, it could be possible for a contrarian zealot to assist in reducing control energy. With conformity behavior, a contrarian zealot always negatively affects network control by increasing energy cost. The results of this paper suggest caution when modelling real networked social systems with the controllability of networked linear dynamics, since the system dynamical behavior is sensitive to parameter change.
\end{abstract}
\maketitle

\begin{quotation}
There has been a lot of interest in studying the controllability of complex networks because many complex systems may be modelled as dynamical systems, thus understanding how to control a high dimensional networked dynamical system has the potential to lead to technological breakthroughs. Typically, in these studies, the type of networked system is not specified, and the system is generically assumed to be linear dynamical. This may pose some problems when one is interested in specifically modelling social dynamics in networks, where the network interactions may have higher complexity. For example, some people may be stubborn (zealots) and refuse to accept new ideas, yet they continue to influence the rest within the social network. This paper addresses this niche by modelling zealots into the framework of network control and investigates how they would affect the energy cost.
\end{quotation}

\section{Introduction}
The controllability of complex networks \cite{liu2011controllability,liu2016control} refers to the modelling of complex dynamical systems with state vector evolving in time and driven by external control signals toward the desired node states. Depending on the system being considered at hand, the state vector represents the amount of traffic which passes through node $i$ in a communication network \cite{pastor2007evolution}, the transcription factor concentration in a gene regulatory network \cite{lezon2006using}, or the opinion of an agent in a consensus network \cite{tanner2004controllability,liu2008controllability,rahmani2009controllability,mesbahi2010graph} and so on. It has its roots in control theory \cite{rugh1996linear,lin1974structural}, and the dynamics of the networked system is assumed to be linear time-invariant (LTI), which is also suitable for modelling opinion networks \cite{tanner2004controllability,liu2008controllability,rahmani2009controllability,mesbahi2010graph}. For modelling complex systems with nonlinear dynamics, such as epidemic spreading in networks \cite{barzel2013universality}, LTI dynamics is adequate for capturing the linearized dynamics of the nonlinear system around its equilibrium points \cite{liu2016control}. Within the literature of network control, an important consideration is the energy cost, which measures the amount of energy that each of the control signal needs to consume to drive the state vector of the network \cite{yan2012controlling}. Therefore, if the energy cost required for performing certain tasks is too high, the system cannot in practice be controlled.   

In statistical physics of social dynamics \cite{castellano2009statistical} (or socio-physics), zealots, agents with unwavering opinions, have been researched in various social dynamics models \cite{mobilia2003does,mobilia2005voting,mobilia2007role,masuda2012evolution,masuda2015opinion,waagen2015effect,verma2014impact,baumann2020laplacian,galam2007role}. For example, if considering a network of opinions on operating systems (such as Microsoft Windows, Apple MAC O/S, or Linux), then a zealot is someone who is fiercely loyal to a particular opinion, who refuses to accept any other views, while advocating theirs to others  \cite{waagen2015effect}. Theoretically, zealots represent interesting modifications to existing models to examine altered system behavior. Empirically, partisanship \cite{evans2018opinion} and confirmation-bias \cite{baumann2020modeling} in social networks have been reported, suggesting that zealots could also be a realistic feature of networked social systems, since not all members are truly neutral. While zealots have become well-understood in the setting of socio-physics, it is still unclear how they affect the behavior of the controllability of complex networks and its associated energy cost. For example, in a campaign to steer the opinions of individuals in a complex social network, would the presence of zealots assist or sabotage the campaigning effort?

In this paper, zealots are introduced to the framework of network controllability, focusing on networked controllability in the context of socio-physics. Conceptually, this research is similar to an earlier work \cite{kafle2018optimal}, which considers two types of drivers, one effecting local influences, and the other the canonical driver nodes, which steers the state vector globally. However, Ref.\ \cite{kafle2018optimal} focuses on the context of infrastructure networks and presents numerical results when the competing driver nodes induce exponentially increasing local influences to simulate infrastructural damages, and examines the amount of energy needed to neutralize these attacks. On the other hand, the present research focuses on competing driver nodes that induce constant local influences, simulating zealots' unwavering opinions, and studies the amount of energy needed to control the social network in competition or cooperation with the zealots. Furthermore, detailed analytical and numerical results are presented, where the number of canonical driver nodes and control time regimes are varied. In addition, beyond the canonical continuous-time linear dynamics \cite{liu2011controllability}, the analyses extend toward discrete-time linear dynamics with conformity behavior \cite{wang2015controlling}, which may be of particular interest to socio-physics, since it models social networks where each agent conforms to their nearest neighbors. Taken together, this paper presents a nuanced characterization of how zealots affect the energy cost when trying to control a complex social network, where it shows that the interplay between different parameters such as the zealots' beliefs, number of drivers, final control time regimes, network effects, network dynamics, number and configurations of nodes influenced by the zealots, can lead to different energy cost behaviors.


\section{Continuous-time linear dynamics model} \label{continuous-time model subsection}
\begin{figure*} 
\begin{center} 
\includegraphics[scale=0.64]{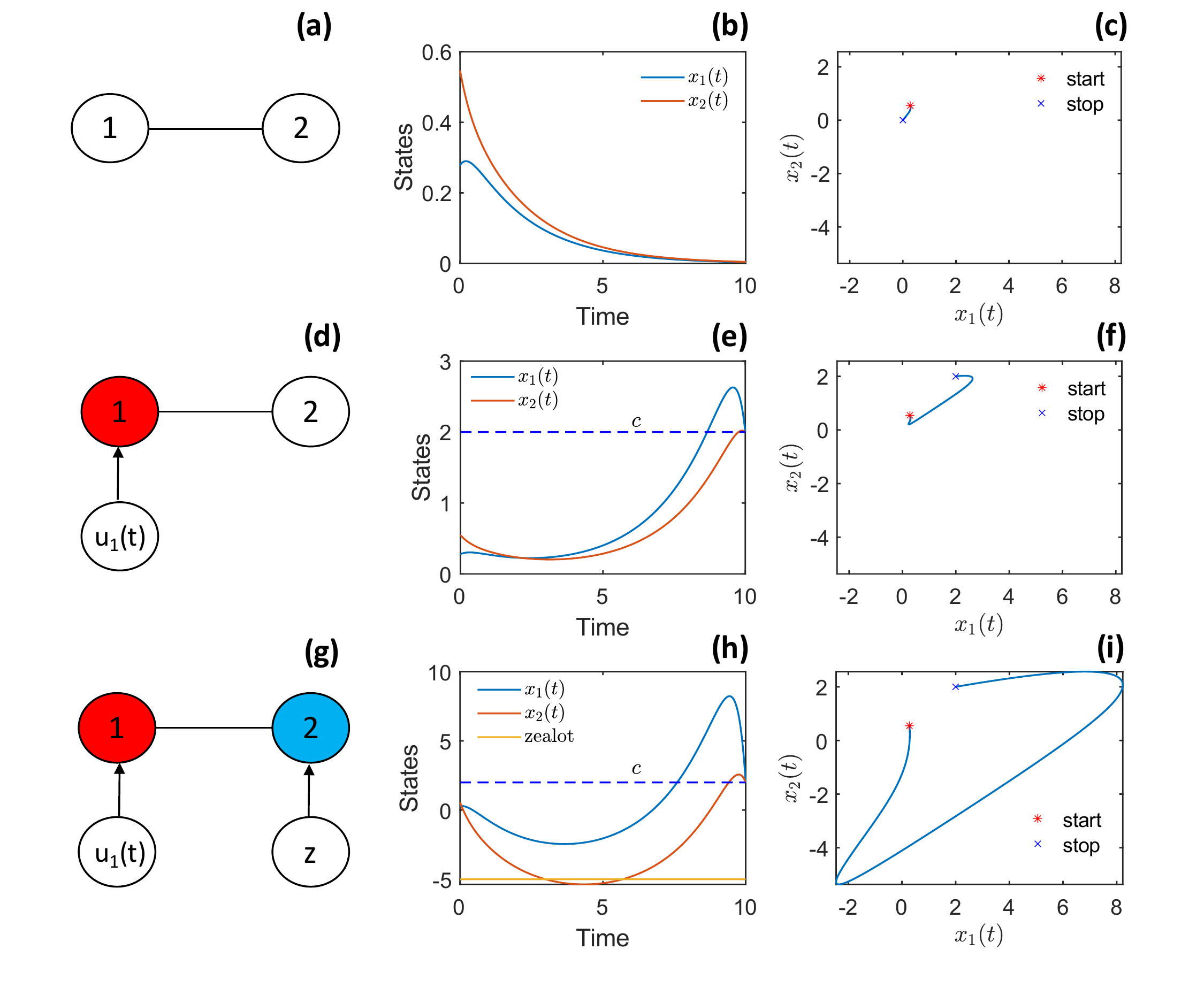}
\end{center}
\caption{(a) Network in the absence of control signal and zealot. (b) Node states evolution of the network in the absence of external influences. (c) State space trajectory of $x_1(t)$ and $x_2(t)$ in the absence of external influences. (d) Network with control signal $u_1(t)$ attached to node $1$. (e) Node states evolution of the network being driven toward consensus $c=2$. (f) State space trajectory of nodes being driven toward consensus. (g) Network with control signal $u_1(t)$ attached to node $1$ and zealot node influencing node $2$. (h) Node states evolution of the network being driven toward consensus, while node $2$ is under the influence of the zealot node with fixed opinion $z=-5$. (i) State space trajectory of nodes being driven toward consensus while under the influence of the zealot. } \label{fig:fig1}
\end{figure*}


To study zealots affecting the energy cost in controlling a complex social network, an example is given in Fig.\ \ref{fig:fig1}. Think of the node states, $x_i(t)$, as opinions on a particular topic, where a positive $x_i(t)$ models support of a particular idea, and a negative $x_i(t)$ denotes opposition. In Fig.\ \ref{fig:fig1}(a), there are $N=2$ number of normal agents (nodes $1$ and $2$), neutral nodes without any preferred opinion, who are open to adopting new ideas, and communicate with one another to exchange information. In Fig.\ \ref{fig:fig1}(d), a control signal $u_1(t)$ attaches to node $1$, making it the driver node \cite{liu2011controllability} (colored red); by directly changing the state of node $1$ with $u_1(t)$, the state of node $2$ becomes affected and all normal agents are controlled toward consensus $c$. In Fig. \ref{fig:fig1}(g), a stubborn agent, or zealot node (node $z$), with fixed opinion $z$, is introduced into the system; the zealot influences node $2$, making it a zealot-influenced node (colored blue). While the driver node is still able to control all normal agents toward consensus $c$, less/more effort may be needed, depending on $c$ and $z$. The states evolution in time are displayed in Fig. \ref{fig:fig1}(b), where they stabilize in long time in the absence of control, and in Figs.\ \ref{fig:fig1}(e) and (h), where the network is driven in the absence and presence of zealots respectively. Correspondingly, the normal agents' state space trajectories are shown in Figs.\ \ref{fig:fig1}(c), (f), and (i). In Fig. \ref{fig:fig1}(i), the $3$rd state dimension would have shown the zealot's fixed node state, which constraints $x_1(t)$ and $x_2(t)$ to the fixed $x_3=z$ plane. For this example, $z=-5$, contrary to the control goal of driving toward $c=2$, and it can be seen in Fig.\ \ref{fig:fig1}(i) that the state space trajectory elongates as compared to Fig.\ \ref{fig:fig1}(f). 

There are a few rules that this model should follow:
\begin{itemize}
\item The mutual exchange of information between normal agents is modelled by undirected links between them 
\item Zealots receive no directed links from any other nodes or control signals as they hold steadfast to their beliefs
\item Zealots advocate their beliefs to other normal agents through directed links
\item External control signal ${\bf u}(t)$ steers the opinions of normal agents with directed links, and a single control signal can only attach to one normal agent.
\end{itemize}

Although self-dynamics links are not shown in the Fig.\ \ref{fig:fig1} models, they should be present for continuous-time network dynamics. Self-dynamics stabilize the dynamics of the system \cite{yan2015spectrum}, which is crucial for modelling complex dynamical systems realistically \cite{cowan2012nodal}, for example, opinion dynamics \cite{acemoglu2010spread}. While there can be multiple zealots present in an arbitrary network, to simplify the scope of the research, only one group of zealots holding the same opinion $z$ is considered, and so it is mathematically the same to model only one zealot node affecting multiple normal agents. Throughout the rest of the paper, ``zealot'', ``zealot node'', or ``zealots'' are used interchangeably. For notational conciseness, the total system size is denoted to be $n$, where $n=N+1$, of which $N$ are normal agents, and the $n$-th node is always the zealot node.

Generalizing to an arbitrary network with $n$ nodes, the continuous-time model with zealots influencing normal agents within the network can be modelled using target control \cite{klickstein2017energy}:
\begin{equation} \label{target control state equation}
\begin{aligned}
&\dot{\bf x}(t) = {\bf A}{\bf x}(t) + {\bf B}{\bf u}(t), \\
&{\bf y}(t) = {\bf C}{\bf x}(t),
\end{aligned}
\end{equation}
where ${\bf x}(t)=[x_1(t),x_2(t),...,x_N(t),z]^T\in \mathbb{R}^{n \times 1}$ is the time-varying state vector, with the first $N$ elements denoting the node states of the $N$ normal agents, and the $n$-th element denoting the zealot node's fixed belief, $z$. ${\bf A}$ is the full $n \times n$ network structure such that $a_{ij}$ is non-zero if there is a directed link from node $j$ to node $i$ ($a_{ij}=0$ otherwise), and comprises the symmetric reduced $N$-dimensional principal submatrix $\tilde{\bf A}$ (remove $n$-th row/column), where non-zero undirected link $\tilde{a}_{ij}=\tilde{a}_{ji}$ represents normal agent nodes $i$ and $j$ that are connected with each other and exchange ideas, with the final row $a_{in}=1$ if zealot node influences node $i$ ($a_{in}=0$ otherwise, and $a_{ni}$ is always zero because the zealot node cannot be influenced by any other nodes). ${\bf B}\in \mathbb{R}^{n \times M}$ is the control input matrix, where $M$ is the number of control signals (such that $1 \leq M \leq N$), and $b_{ij}=1$ if control signal $j$ attaches to node $i$ (for $i=1,2,...,N$, where $i=n$ is not permitted as the zealot node has unwavering opinion, and $j=1,2,...,M$). Nodes which have a control signal attached to them are called driver nodes. ${\bf u}(t)=[u_1(t),u_2(t),...,u_M(t)]^T \in \mathbb{R}^{M \times 1}$ is the input vector of external control signals. ${\bf y}(t)=[y_1(t),y_2(t),...,y_N(t)]^T \in \mathbb{R}^{N \times 1}$ is the output state vector. ${\bf C} \in \mathbb{R}^{N \times n}$ is the target control matrix which relates which node states are being steered by ${\bf u}(t)$, and $c_{ij}=1$ if node $j$ is the $i$-th node (where $i,j=1,2,...,N$) to be target controlled, and $c_{ij}=0$ otherwise. While Eqn.\ (\ref{target control state equation}) borrows the language of target control \cite{klickstein2017energy}, note that for this research, full controllability of all $N$ normal agents are considered, thus identity ${\bf I}_N$ is the reduced $N$-dimensional principal submatrix of ${\bf C}$, such that only the first $N$ nodes of the state vector is being driven (final column $c_{in}=0$ for $i=1,2,...,N$ since the zealot node has fixed opinion, $z$, controlling the $n$-th node is not permitted).

The energy cost is defined to be \cite{rugh1996linear} 
\begin{equation} \label{cost function}
J=\int_{t_0}^{t_f} {\bf u}^T(t){\bf u}(t) dt,
\end{equation}
where $t_0$ is the initial time, $t_f$ is the final control time (the amount of time allocated to the control signals to steer the state vector), and when minimized leads to the energy-optimal target control signal \cite{klickstein2017energy}
\begin{equation} \label{optimal target control signal}
{\bf u}^{*}(t)={\bf B}^T e^{{\bf A}^T(t_f-t)}{\bf C}^T({\bf C}{\bf W} {\bf C}^T)^{-1}({\bf y}_f-{\bf C}e^{{\bf A}(t_f-t_0)}{\bf x}_0),
\end{equation}
where ${\bf W}=\int_{t_0}^{t_f}e^{{\bf A}(t_f-t)}{\bf B}{\bf B}^T e^{{\bf A}^T(t_f-t)} dt$ is the controllability Gramian, ${\bf y}_f=[c,c,...,c]^T \in \mathbb{R}^{N \times 1}$ is the final output state vector where $c$ is the consensus opinion that the system is being steered toward, and ${\bf x}_0=[0,0,...,0,z]^T \in \mathbb{R}^{n \times 1}$ is the initial state vector of the system, where it is assumed that all node states begin initially with neutral opinions at zero. When ${\bf B}$ matrix is chosen appropriately such that the system is controllable, $({\bf C}{\bf W}{\bf C}^T)$ is invertible \cite{klickstein2017energy}. Substituting Eqn.\ (\ref{optimal target control signal}) into Eqn.\ (\ref{cost function}), and setting $t_0=0$, the energy cost when using ${\bf u}^*(t)$ to steer the complex system is
\begin{equation}\label{energy-optimal energy cost}
\begin{aligned}
\mathcal{E}=& {\bf y}_f^T ({\bf C}{\bf W}{\bf C}^T)^{-1}{\bf y}_f -2{\bf x}_0^T e^{{\bf A}^T t_f} {\bf C}^T ({\bf C} {\bf W} {\bf C}^T )^{-1} {\bf y}_f \\
&+ {\bf x}_0^T e^{{\bf A}^T t_f}{\bf C}^T({\bf C}{\bf W}{\bf C}^T)^{-1}{\bf C}e^{{\bf A}t_f}{\bf x}_0
\end{aligned}
\end{equation}
(for derivation of Eqn.\ (\ref{energy-optimal energy cost}) and subsequent derivations, see Supplementary Information). 

The reduced network connection matrix, $\tilde{\bf A}$, which consists of only normal agents' mutual interactions (undirected links), is symmetric and can be diagonalized as $\tilde{\bf A}=\tilde{\bf P}\tilde{\bf D}\tilde{\bf P}^T$, where $\tilde{\bf P}$ is the $N \times N$ orthonormal eigenvectors matrix such that $\tilde{\bf P}\tilde{\bf P}^T=\tilde{\bf P}^T\tilde{\bf P}={\bf I}$, and $\tilde{\bf D}=\text{diag}\{\lambda_1,\lambda_2,...,\lambda_N\}$ is a diagonal matrix containing the eigenvalues of $\tilde{\bf A}$, where they are ordered ascendingly: $\lambda_1 \leq \lambda_2 \leq ... \leq \lambda_N$. Equivalently, $\tilde{\bf A}$ is obtained by removing the $n$-th row/column off the full network matrix, ${\bf A}$. The full network matrix, $\bf A$, includes the directed links of the zealot node to normal agents in the last column, $a_{in}$, is non-symmetric and thus eigen-decomposed as $n\times n$ dimensional matrices ${\bf A}={\bf P}{\bf D}{\bf P}^{-1}$ and ${\bf A}^T={\bf V}{\bf D}{\bf V}^{-1}$, where ${\bf P}$ and ${\bf V}$ are the eigenvectors matrices of ${\bf A}$ and ${\bf A}^T$ respectively, and ${\bf D}=\text{diag}\{\lambda_1,\lambda_2,...,\lambda_N,0\}$ comprises $\tilde{\bf D}$ as its reduced $N$-dimensional principal submatrix, with the $n$-th diagonal entry being $0$ (zealot node does not have self-link since it has fixed $z$ opinion). Most complex systems tend to operate near a stable state, so for the continuous-time model, the eigenvalues $\lambda_i$ (for $i=1,2,...,N$) are all negative \cite{may2019stability,yan2015spectrum}, and network $\tilde{\bf A}$ is negative definite (ND). Note that matrix terms containing $\sim$ symbols refer to the $N\times N$ reduced matrix without the zealot node, while those without refer to the $n \times n$ full matrix.

For controlling all $N$ normal agents, $({\bf C}{\bf W}{\bf C}^T)=\tilde{\bf W}$, where $\tilde{\bf W}=\int_{0}^{t_f}e^{\tilde{\bf A}(t_f-t)}\tilde{\bf B}\tilde{\bf B}^T e^{\tilde{\bf A}^T(t_f-t)}dt$
is the controllability Gramian of the reduced network connection matrix, $\tilde{ \bf A}$, and $\tilde{\bf B}\in \mathbb{R}^{N\times M}$ is the input control matrix of the reduced network matrix, with $\tilde{b}_{ij}=1$ if control signal $j$ attaches to node $i$ ($\tilde{b}_{ij}=0$ otherwise). The controllability Gramian, $\tilde{\bf W}$, can be expressed as the Hadamard product analytical form \cite{yan2012controlling,yan2015spectrum,duan2019energy} 
\begin{equation}
\begin{aligned}
\tilde{\bf W}= & \tilde{\bf P}\tilde{\bf M}\tilde{\bf P}^T = & \tilde{\bf P}[\tilde{\bf Q} \circ\tilde{\bf F}]\tilde{\bf P}^T, 
\end{aligned}
\end{equation}
where $\tilde{\bf M}=\tilde{\bf Q} \circ\tilde{\bf F}$ is the simplified controllability Gramian of the reduced matrix, and
\begin{equation} \label{M_ij Q_ij}
\begin{aligned}
\tilde{\bf M}_{ij}=&\tilde{\bf Q}_{ij}\tilde{\bf F}_{ij}=&[\tilde{\bf P}^T\tilde{\bf B}\tilde{\bf B}^T\tilde{\bf P}]_{ij}\frac{[e^{(\lambda_i+\lambda_j)t_f}-1]}{\lambda_i+\lambda_j},
\end{aligned}
\end{equation} 
with $\tilde{\bf Q}_{ij}=[\tilde{\bf P}^T\tilde{\bf B}\tilde{\bf B}^T\tilde{\bf P}]_{ij}$, and $\tilde{\bf F}_{ij}=\frac{[e^{(\lambda_i+\lambda_j)t_f}-1]}{\lambda_i+\lambda_j}$. The inverse matrices follow similarly as 
\begin{equation} \label{M inverse}
({\bf C}{\bf W}{\bf C}^T)^{-1}=\tilde{\bf W}^{-1}=(\tilde{\bf P} \tilde{\bf M} \tilde{\bf P}^T)^{-1}=\tilde{\bf P}\tilde{\bf M}^{-1}\tilde{\bf P}^T.
\end{equation}

Substituting the eigen-decompositions ${\bf A}={\bf P}{\bf D}{\bf P}^{-1}$, ${\bf A}^T={\bf V}{\bf D}{\bf V}^{-1}$, $\tilde{\bf A}=\tilde{\bf P}\tilde{\bf D}\tilde{\bf P}^T$, and Eqn.\ (\ref{M inverse}) into Eqn.\ (\ref{energy-optimal energy cost}), the energy cost becomes
\begin{equation} \label{energy cost with quadratic z form} 
\begin{aligned}
\mathcal{E}=&c^2 \sum_{i=1}^N \sum_{j=1}^N[\tilde{\bf P}\tilde{\bf M}^{-1}\tilde{\bf P}^T]_{ij}\\
&-2cz\sum_{i=1}^N\sum_{j=1}^N[{\bf V}e^{{\bf D}t_f}{\bf V}^{-1}]_{ni}[\tilde{\bf P}\tilde{\bf M}^{-1}\tilde{\bf P}^T]_{ij}\\
&+z^2 \sum_{i=1}^N \sum_{j=1}^N [{\bf V}e^{{\bf D}t_f}{\bf V}^{-1}]_{ni}[\tilde{\bf P}\tilde{\bf M}^{-1}\tilde{\bf P}^T]_{ij}[{\bf P}e^{{\bf D}t_f}{\bf P}^{-1}]_{jn}.
\end{aligned}
\end{equation}
By inspection, Eqn.\ (\ref{energy cost with quadratic z form}) is a quadratic function in terms of $z$, given that $c$ is fixed. Solving the turning point $\frac{\partial \mathcal{E}}{\partial z}=0$, $z^*$ can be obtained, where $z^*$ is the opinion of the zealot node which yields the lowest energy cost when controlling the system toward consensus $c$. Further, the network structure $\tilde{\bf A}$, as well as the nodes which are influenced by the zealot node can also cause a change in the control energy. 

\subsection{Analytical equations of energy cost}
Depending on the final control time, $t_f$, and driver nodes placement, $\tilde{\bf M}^{-1}$ changes and Eqn.\ (\ref{energy cost with quadratic z form}) changes accordingly, leading to different behaviors. Subsequently, similar to Refs.\ \cite{yan2012controlling} and \cite{duan2019energy}, the energy cost is analyzed in terms of differing number of driver nodes through control input matrix $\tilde{\bf B}$, which is encoded in $\tilde{\bf M}$, and different final control time regimes (small $t_f$ and large $t_f$). 

Controlling the system with $N$ driver nodes such that all $N$ normal agents each receive a control signal, $\tilde{\bf B}={\bf I}_N$, $\tilde{\bf Q} = {\bf I}_N$, and $\tilde{\bf M}$ becomes a diagonal matrix with its main diagonal entries being $\tilde{\bf M}_{ii}=\tilde{\bf F}_{ii}=\frac{1}{2\lambda_i}[e^{2\lambda_it_f}-1]$. $\tilde{\bf M}^{-1}$ is also a diagonal matrix, with $[\tilde{\bf M}^{-1}]_{ii}=\frac{1}{\tilde{\bf M}}_{ii}=\frac{2\lambda_i}{[e^{2\lambda_i t_f}-1]}$, which in the small $t_f$ limit, using Taylor expansion $e^{2\lambda_i t_f}\approx 1+2\lambda_i t_f$, and in the large $t_f$ limit, as the eigenvalues are negative $[e^{2\lambda_i t_f}-1]\approx -1$, leading to
\begin{equation}\label{N driver piecewise M inverse}
\tilde{\bf M}^{-1}(i,i)\approx \begin{cases}
-2\lambda_i, & \text{large }t_f,\\
t_f^{-1}, & \text{small }t_f.
\end{cases}
\end{equation}

When controlling the system with one driver node, where there is a single control signal $u_1(t)$ attached to arbitrary node $h$, then $\tilde{\bf B}$ is a $N \times 1$ matrix with $\tilde{b}_{h1}=1$, leading to $\tilde{\bf B}\tilde{\bf B}^T=\tilde{\bf J}^{hh}$, where $\tilde{\bf J}^{hh}$ is a $N \times N$ single-entry matrix \cite{petersen2012matrix} such that $[\tilde{\bf J}^{hh}]_{ij}=1$ when $i=j=h$, and zero otherwise. Consequently, $\tilde{\bf Q}_{ij}=\tilde{p}_{hi}\tilde{p}_{hj}$, $\tilde{\bf M}_{ij}=\tilde{p}_{hi}\tilde{p}_{hj}[\frac{e^{(\lambda_i + \lambda_j)t_f}-1}{\lambda_i+\lambda_j}]$, and in the large $t_f$ limit, the exponential terms vanish because all $\lambda_i$ are negative, and $\tilde{\bf M}_{ij}=\frac{-\tilde{p}_{hi}\tilde{p}_{hj}}{\lambda_i+\lambda_j}$. Using $\tilde{\bf M}^{-1}=\frac{\tilde{\bf M}^*}{|\tilde{\bf M}|}$, where $\tilde{\bf M}^*$ and $|\tilde{\bf M}|$ are the adjoint matrix and determinant of $\tilde{\bf M}$ respectively, \cite{duan2019energy}
\begin{equation}\label{M inverse one driver large tf ND}
\tilde{\bf M}^{-1}(i,j)= \frac{-4\lambda_i \lambda_j}{\tilde{p}_{hi}\tilde{p}_{hj}(\lambda_i+\lambda_j)}\prod_{\substack{k=1\\k\neq i}}^{N}\frac{\lambda_i+\lambda_k}{\lambda_i-\lambda_k} \prod_{\substack{k=1 \\ k \neq j}}^{N}\frac{\lambda_j+\lambda_k}{\lambda_j-\lambda_k}.
\end{equation}
For small $t_f$, neither first-order nor second-order Taylor expansion of $\tilde{\bf F}_{ij}$ yields an invertible $\tilde{\bf M}$, and $\tilde{\bf M}^{-1}(i,j)$ has to be estimated with \cite{duan2019energy} $|\tilde{\bf M}|\sim t_f^{N_0}$ and $\tilde{\bf M}^*(i,j)\sim t_f^{N_{ij}}$ such that
\begin{equation}\label{small t_f M inverse}
\tilde{\bf M}^{-1}(i,j)\sim t_f^{N_{ij}-N_0},
\end{equation}
where the integer exponents $N_{ij}$ (for $i,j=1,2,...,N$) or $N_0$ refer to the $N_{ij}$-th or $N_0$-th order Taylor expansion of $\tilde{\bf F}_{ij}$ where invertibility is satisfied, and are computed numerically. 

Using $d$ number of drivers to control the network, where $1<d<N$, $\tilde{\bf B}$ is a $N \times d$ matrix with $\tilde{b}_{ij}=1$ if control signal $j$ (where $j=1, 2, ..., d$) attaches to node $i$ (where $i=1, 2, ..., N$), leading to $\tilde{\bf B}\tilde{\bf B}^T=\sum\limits_{k=1}^{d}{\bf J}^{d_k d_k}$, where $d_k=\{1,2,...,N\}$ refers to the arbitrary $k$-th driver node. Thereafter, $\tilde{\bf Q}_{ij}=\sum\limits_{k=1}^{d}\tilde{p}_{d_k i}\tilde{p}_{d_k j}$, and $\tilde{\bf M}_{ij}=\sum\limits_{k=1}^{d}\tilde{p}_{d_k i}\tilde{p}_{d_k j}\frac{[e^{(\lambda_i+\lambda_j)t_f}-1]}{\lambda_i+\lambda_j}$. Owing to the summation, it is difficult factor the terms to derive $\tilde{\bf M}^{-1}$ analytically. Thus, $d$ drivers energy cost results have to be computed with numerical $\tilde{\bf M}^{-1}$.

Substituting Eqs.\ (\ref{N driver piecewise M inverse}) and (\ref{M inverse one driver large tf ND}) into Eqn.\ (\ref{energy cost with quadratic z form}), the analytical energy costs are (the superscripts denote the $t_f$ regime, and the subscripts denote the number of drivers used to control the network) 

\begin{equation}\label{E_N large tf}
\begin{aligned}
\mathcal{E}^{\text{large }t_f}_{N} = &  -2c^2 \sum_{i,j,k} \tilde{p}_{ik}\tilde{p}_{jk}\lambda_k +4cz \sum_{i,j,k} [{\bf V}^{-1}]_{ni}\tilde{p}_{ik}\tilde{p}_{jk}\lambda_k\\
&-2z^2 [{\bf P}^{-1}]_{nn} \sum_{i,j,k} [{\bf V}^{-1}]_{ni} \tilde{p}_{ik}\tilde{p}_{jk} p_{jn} \lambda_k,
\end{aligned}
\end{equation}
where $\sum\limits_{i,j,k}=\sum\limits_{i=1}^{N}\sum\limits_{j=1}^{N}\sum\limits_{k=1}^{N}$,
\begin{equation}\label{E_N small tf}
\mathcal{E}^{\text{small }t_f}_{N} \approx c^2 t_f^{-1} N - 2czr + z^2 r t_f, 
\end{equation}
where $r=\sum\limits_{i=1}^N a_{in}$ is the integer number of nodes which are influenced by the zealot node (the $n$-th node),
\begin{equation}\label{E_1 large tf} 
\begin{aligned} 
&\mathcal{E}_{1}^{\text{large }t_f} =  \\
& c^2 \sum_{i,j,l,m} \tilde{p}_{li} \tilde{p}_{mj} \frac{-4\lambda_i \lambda_j}{\tilde{p}_{hi}\tilde{p}_{hj}(\lambda_i+\lambda_j)}\prod_{\substack{k=1\\k\neq i}}^{N}\frac{\lambda_i+\lambda_k}{\lambda_i-\lambda_k} \prod_{\substack{k=1 \\ k \neq j}}^{N}\frac{\lambda_j+\lambda_k}{\lambda_j-\lambda_k}\\
&-2cz \sum_{i,j,l,m} [{\bf V}^{-1}]_{nl} \tilde{p}_{li} \tilde{p}_{mj} \frac{-4\lambda_i \lambda_j}{\tilde{p}_{hi}\tilde{p}_{hj}(\lambda_i+\lambda_j)}\prod_{\substack{k=1\\k\neq i}}^{N}\frac{\lambda_i+\lambda_k}{\lambda_i-\lambda_k} \prod_{\substack{k=1 \\ k \neq j}}^{N}\frac{\lambda_j+\lambda_k}{\lambda_j-\lambda_k}\\
&+ z^2 \sum_{i,j,l,m} [{\bf V}^{-1}]_{nl}\tilde{p}_{li}\tilde{p}_{mj}p_{mn} [{\bf P}^{-1}]_{nn} \frac{-4\lambda_i \lambda_j}{\tilde{p}_{hi}\tilde{p}_{hj}(\lambda_i+\lambda_j)}\prod_{\substack{k=1\\k\neq i}}^{N}\frac{\lambda_i+\lambda_k}{\lambda_i-\lambda_k} \\
& \times \prod_{\substack{k=1 \\ k \neq j}}^{N}\frac{\lambda_j+\lambda_k}{\lambda_j-\lambda_k},
\end{aligned}
\end{equation}
where $\sum\limits_{i,j,l,m}=\sum\limits_{i=1}^N \sum\limits_{j=1}^N \sum\limits_{l=1}^N \sum\limits_{m=1}^N$. 

For small $t_f$ regime, one driver energy cost, although Eqn.\ (\ref{small t_f M inverse}) is a valid approximation, owing to the coupling terms $[{\bf V}e^{{\bf D}t_f}{\bf V}^{-1}]_{ni}[\tilde{\bf P}\tilde{\bf M}^{-1} \tilde{\bf P}^T]_{ij}$ and $[{\bf V}e^{{\bf D}t_f}{\bf V}^{-1}]_{ni}[\tilde{\bf P}\tilde{\bf M}^{-1} \tilde{\bf P}^T]_{ij}[{\bf P}e^{{\bf D}t_f}{\bf P}^{-1}]_{jn}$ in Eqn.\ (\ref{energy cost with quadratic z form}), it is difficult to express small $t_f$ $\tilde{\bf M}^{-1}(i,j)$ terms in approximate form. Therefore, numerical $\tilde{\bf M}^{-1}$ is used instead. Furthermore, since $d$ drivers energy cost also require numerical $\tilde{\bf M}^{-1}$, all three analytical energy cost equations should thus be expressed by Eqn.\ (\ref{energy cost with quadratic z form}). Letting ${\bf V}e^{{\bf D}t_f} {\bf V}^{-1}= e^{{\bf A}^T t_f}$, and ${\bf P}e^{{\bf D}t_f} {\bf P}^{-1}= e^{{\bf A} t_f}$,

\begin{equation} \label{E_1 small t_f and E_d} 
\begin{aligned}
\begin{rcases}
\mathcal{E}_1^{\text{small }t_f}\\
\mathcal{E}_d^{\text{large }t_f} \\
\mathcal{E}_d^{\text{small }t_f}
\end{rcases}
= &c^2 \sum_{i=1}^N \sum_{j=1}^N [\tilde{\bf P}\tilde{\bf M}^{-1} \tilde{\bf P}^T]_{ij}\\
& - 2cz\sum_{i=1}^N \sum_{j=1}^N [e^{{\bf A}^T t_f}]_{ni} [\tilde{\bf P}\tilde{\bf M}^{-1} \tilde{\bf P}^T]_{ij}\\
& + z^2  \sum_{i=1}^N \sum_{j=1}^N [e^{{\bf A}^T t_f}]_{ni}[\tilde{\bf P}\tilde{\bf M}^{-1} \tilde{\bf P}^T]_{ij}[e^{{\bf A}t_f}]_{jn},
\end{aligned}
\end{equation}
where the driver nodes placement are encoded in $\tilde{\bf M}^{-1}$ through the inverse of Eqn.\ (\ref{M_ij Q_ij}), and the time regimes are set by $t_f$.

From the presented analytical energy cost equations, note that the choice of zealot-influenced nodes affects the energy cost. In the large $t_f$ regime, such as Eqns.\ (\ref{E_N large tf}) and (\ref{E_1 large tf}), the choice of zealot-influenced nodes enters the equations through $[{\bf V}^{-1}]_{ni}$ and $p_{jn}$, which are respectively the final row and final column of the eigenvectors matrix of the full (transposed) network, ${\bf A}^T$ and ${\bf A}$. Depending on which nodes are being influenced by the zealot node (through $a_{in}=\{0,1\}$), $[{\bf V}^{-1}]_{ni}$ and $p_{jn}$ change accordingly, and the choice of influenced nodes is consequential to the energy cost. One the other hand, in the small $t_f$ regime, the $N$ drivers energy cost is invariant to the choice of zealot-influenced nodes. In Eqn.\ (\ref{E_N small tf}), ceteris paribus, only the $r$ number (and not choice) of zealot-influenced nodes tunes the energy cost.

Taking $\frac{\partial \mathcal{E}}{\partial z}=0$, the minima are (subscripts denote number of drivers and $t_f$ regime)
\begin{equation} \label{z minimum N large tf}
z^*_{N,\text{ large }t_f} =\frac{c\sum\limits_{i,j,k}[{\bf V}^{-1}]_{ni}\tilde{p}_{ik}\tilde{p}_{jk}\lambda_k}{[{\bf P}^{-1}]_{nn}\sum\limits_{i,j,k}[{\bf V}^{-1}]_{ni}\tilde{p}_{ik}\tilde{p}_{jk}p_{jn} \lambda_k},
\end{equation}
\begin{equation} \label{z minimum N small tf}
z^*_{N,\text{ small }t_f}\approx c t_f^{-1},
\end{equation}
\begin{equation} \label{z minimum 1 large tf}  \footnotesize
\begin{aligned}
& z^*_{1, \text{ large }t_f}=\\
& \frac{c \sum\limits_{i,j,l,m}[{\bf V}^{-1}]_{nl}\tilde{p}_{li}\tilde{p}_{mj} \frac{-4\lambda_i \lambda_j}{\tilde{p}_{hi}\tilde{p}_{hj}(\lambda_i + \lambda_j)}\prod\limits_{\substack{k=1\\k\neq i}}^N \frac{\lambda_i + \lambda_k}{\lambda_i - \lambda_k} \prod\limits_{\substack{k=1\\k\neq i}}^N \frac{\lambda_i + \lambda_k}{\lambda_i - \lambda_k}}{\sum\limits_{i,j,l,m}[{\bf V}^{-1}]_{nl}\tilde{p}_{li}\tilde{p}_{mj}p_{mn}[{\bf P}^{-1}]_{nn}\frac{-4\lambda_i \lambda_j}{\tilde{p}_{hi}\tilde{p}_{hj}(\lambda_i + \lambda_j)}\prod\limits_{\substack{k=1\\k\neq i}}^N \frac{\lambda_i + \lambda_k}{\lambda_i - \lambda_k} \prod\limits_{\substack{k=1\\k\neq i}}^N \frac{\lambda_i + \lambda_k}{\lambda_i - \lambda_k}},
\end{aligned}
\end{equation}
and

\begin{equation} \label{z minimum 1 driver small tf, and d drivers}
\begin{rcases}
z^*_{1, \text{ small }t_f}\\
z^*_{d, \text{ large }t_f} \\
z^*_{d, \text{ small }t_f}
\end{rcases}
= \frac{c \sum\limits_{i=1}^{N}\sum\limits_{j=1}^{N} [e^{{\bf A}^T t_f}]_{ni} [\tilde{\bf P} \tilde{\bf M}^{-1}\tilde{\bf P}^T]_{i j} }{\sum\limits_{i=1}^{N} \sum\limits_{j=1}^{N} [e^{{\bf A}^T t_f}]_{ni}[\tilde{\bf P} \tilde{\bf M}^{-1}\tilde{\bf P}^T]_{ij} [e^{{\bf A}t_f}]_{jn}}.
\end{equation}

When the zealots hold unwavering optimal opinion $z^*$, their presence assist in driving the network opinions toward $c$ by lowering the energy cost, as compared to the situation where there were no zealots present. In the small $t_f$ regime, when using $N$ drivers to control the complex network, the optimal zealot $z^*_{N,\text{ small }t_f}$ opinion, is unaffected by network properties, and is proportional to consensus $c$ and $t_f^{-1}$. In contrast, all other turning points $z^*$ are affected by network properties. In those cases, it is difficult to analyze $z^*$ by inspection owing to the many coupled terms present, and numerical experiments are needed to gain more insight. 

\subsection{Numerical experiments}

\begin{figure*} 
\begin{center} 
\includegraphics[scale=0.64]{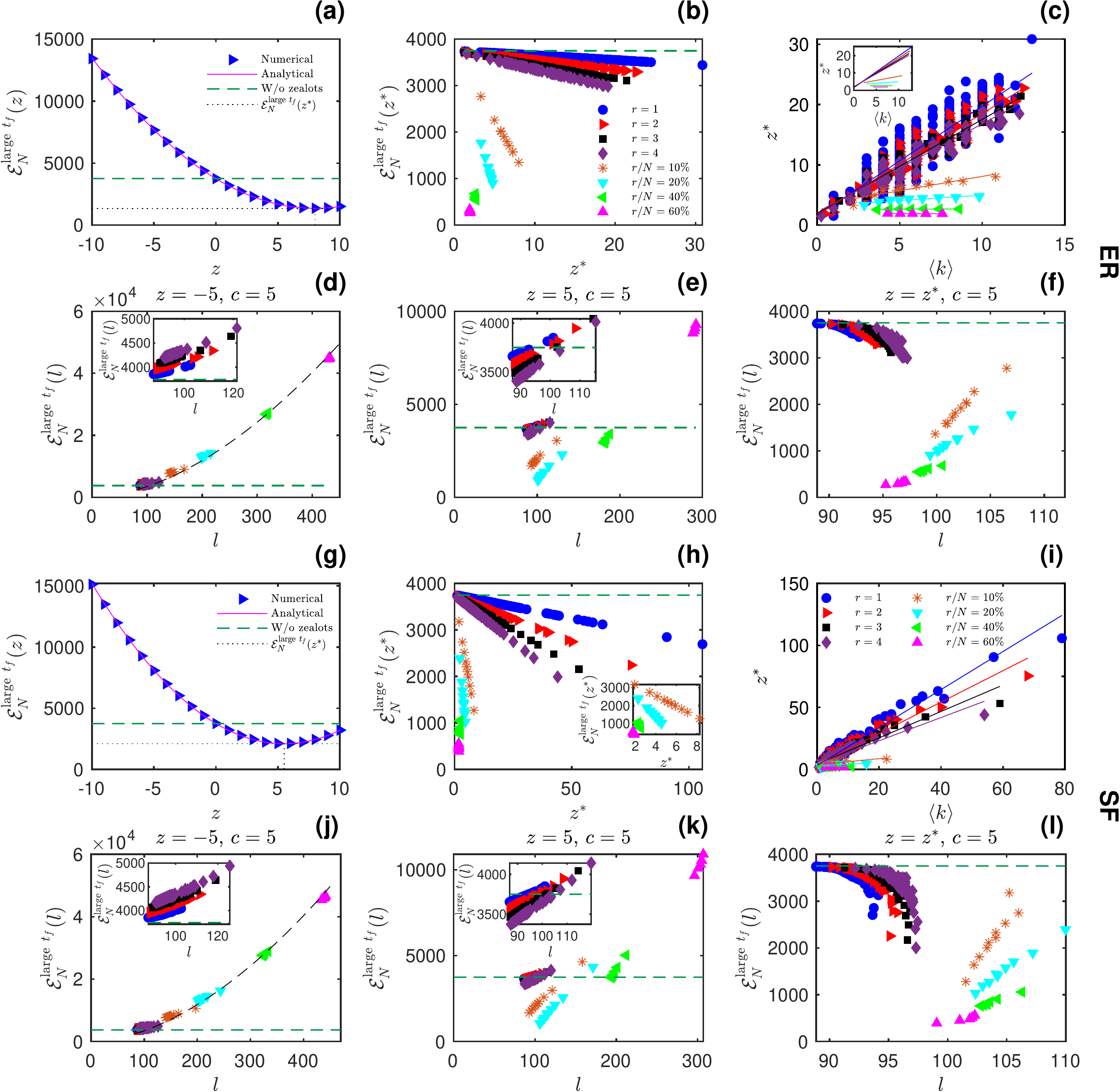}
\end{center}
\caption{Continuous-time linear dynamics: Large $t_f$ regime, $N$ drivers results in ER and SF networks with varying configurations of zealot-influenced nodes. (a)\textemdash(f) are results from a ER network, while (g)\textemdash(l) are results from a SF network. (a) and (g) validate Eqn.\ (\ref{E_N large tf}). (b) and (h) show that as the turning points $z^*$ increases, their associated minima energy costs $\mathcal{E}^{\text{large }t_f}_N(z^*)$ decreases. (c) and (i) show that the average node degree $\langle k \rangle$ of a configuration of zealot-influenced nodes correlate to its turning point $z^*$. Inset in (c) show the linear fits of all the data points. (d) and (j) show that length $l$ is proportional to energy cost for different configurations of nodes influenced by a zealot of fixed $z=-5$ opinion (black dashed lines in both are fitted by $\mathcal{E}^{\text{large }t_f}_N(l)\approx l^{1.77}$). (e) and (k) are the same, except that the zealot has fixed opinion $z=5$. Likewise, (f) and (l) are the same, except that the zealot has fixed opinion $z=z^*$. The insets in (d), (e), (j), and (k) show the data points associated with $r=1$, $r=2$, $r=3$, $r=4$ which are clustered closely. The different markers denote the varying $r$ number of zealot-influenced normal agents according to legend in (b) or (i), and green dashed line indicates the energy cost in the absence of any zealot.} \label{fig:fig2}
\end{figure*}

\begin{figure*} 
\begin{center} 
\includegraphics[scale=0.64]{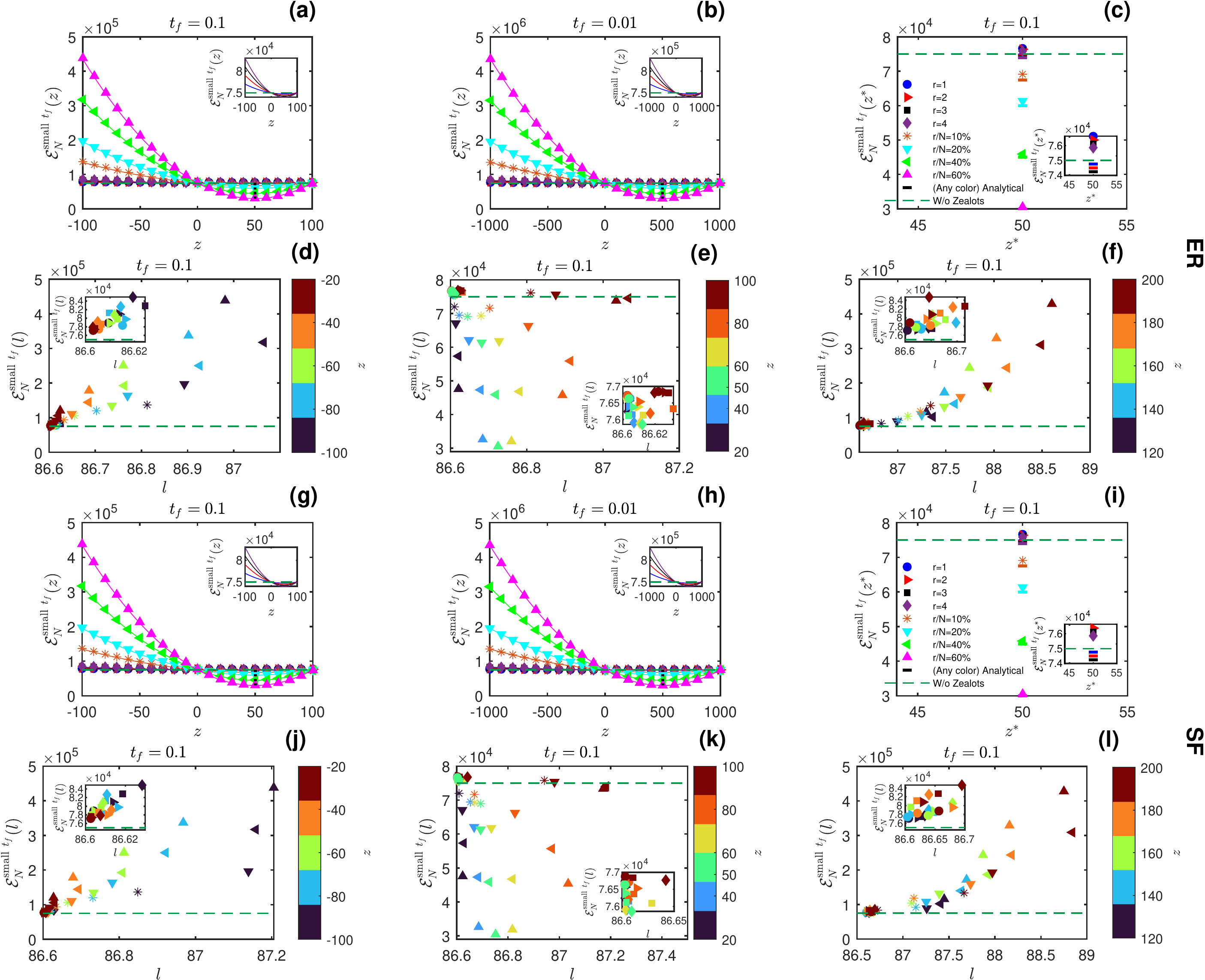}
\end{center}
\caption{Continuous-time linear dynamics: Small $t_f$ regime, $N$ drivers results in ER and SF networks. Unlike the large $tf$ regime results, varying the choice of zealot-influenced normal agents is inconsequential to the energy cost, and only the $r$ number of zealot-influenced normal agents matters. (a)\textemdash(f) are results from a ER network, while (g)\textemdash(l) are results from a SF network. (a) and (g) ((b) and (h)) validate Eqn.\ (\ref{E_N small tf}) at $t_f=0.1$ ($t_f=0.01$) for different $r$ values, where the insets show the analytical energy cost curves associated with $r=1$, $r=2$, $r=3$, and $r=4$. Note that the turning points $z^*$ is invariant to the $r$ number of zealot-influenced nodes. (c) and (i) show the turning points $z^*$ and associated minima energy costs (insets show the data points associated with $r=1$, $r=2$, $r=3$, $r=4$, where there are slight deviations between analytical and numerical results due to the first-order approximation $e^{{\bf A}t_f}\approx {\bf I}_{n} + {\bf A}t_f$ used in deriving analytical results). (d) and (j) display the energy cost and associated state space trajectory length $l$ when different $r$ number of zealot-influenced nodes is varied (the symbol markers correspond to $r$ value according to legends in (c) or (i)), as well as the opinion $z$ of the zealot is varied from $[-100,-20]$ (color represent the associated $z$ values). (e) and (k) are the same, except that $z\in [20,100]$. Likewise, (f) and (l) are the same, except that $z\in [120,200]$. Green dashed line indicates the energy cost in the absence of any zealot. } \label{fig:fig3}
\end{figure*}

In the results that follows, it can be assumed that $t_f=20$ and $t_f\leq 0.1$ respectively for large and small $t_f$ regimes, and the complex networks are being driven toward consensus $c=5$, which is fixed throughout, while $z$, the zealot's unwavering opinion is tuned, and the choice of nodes influenced by the zealot node (normal agent nodes which receive directed links from the zealot node) is varied.


\subsubsection{N drivers}

The results of the numerical experiments when controlling random Erdős–Rényi \cite{albert2002statistical} (ER) and scale-free (SF; static model \cite{goh2001universal}) complex networks topologies toward consensus $c=5$,  with $N=300$ normal agents, and average degree $\langle k \rangle=6$, using $N$ number of drivers, such that each node directly receives a control signal $u_i(t)$ are presented in Figs.\ \ref{fig:fig2} and \ref{fig:fig3} for large $t_f$ and small $t_f$ respectively. 

As expected, the analytical energy cost $\mathcal{E}(z)$, as a function of varying fixed zealot opinion $z$ (Eqns.\ (\ref{E_N large tf}) and (\ref{E_N small tf})), validated against numerical computations, show a quadratic behavior with respect to $z$ in Figs.\ \ref{fig:fig2}(a), \ref{fig:fig2}(g), \ref{fig:fig3}(a), \ref{fig:fig3}(b), \ref{fig:fig3}(g), and \ref{fig:fig3}(h). Therefore, depending on what the zealot's fixed opinion $z$ is, the energy cost needed to control a complex network to consensus $c=5$ follows a quadratic curve, which has a turning point at $z^*$ (black dotted line) that assists in lowering the energy cost, as compared to the energy cost needed for controlling the complex network in a situation where there are no zealots present, denoted by the horizontal green dashed line intersecting with the quadratic curve at $z=0$. Away from minima $z^*$, the energy cost increases at a $z^2$ rate, which when above the green dashed line, opinion $z$ becomes detrimental to the controlling of complex networks, and the zealots' presence increases energy cost. 


For the large $t_f$ regime results, Figs.\ \ref{fig:fig2}(a) and (g) correspond to the energy costs (Eqn.\ (\ref{E_N large tf})) needed to control ER and SF networks when a particular set of nodes have been influenced by the zealot node, with strength of zealot opinion $z$ varying in the range $[-10,10]$. In any one instance of a numerical experiment, the zealot node can influence $r$ number of normal agent nodes, which are fixed, once chosen, for the entirety of that instance. Depending on the configuration of zealot-influenced nodes, the parabolas as shown in Figs.\ \ref{fig:fig2}(a) and (g) vary accordingly with different turning points $z^*$, and their associated minimum energy costs $\mathcal{E}^{\text{large }t_f}_N(z^*)$, and steepness. To understand how the selection of $r$ number of zealot-influenced nodes affects the energy cost, their configurations were varied with increasing $r$, from $r=1$, $r=2$, $r=3$, $r=4$, up to $r/N=60\%$ of the network nodes, and the turning points $z^*$, minima energy costs $\mathcal{E}^{\text{large }t_f}_N(z^*)$, and associated state space trajectory lengths were measured.

The state space trajectory was first introduced in Fig.\ \ref{fig:fig1}, for example, when the zealot node (holding fixed zealot opinion $z=-5$) is introduced into the system (for controlling network node states toward consensus $c=2$), the length of the state space trajectory elongates. Mathematically, the length $l$ of the state space trajectory is computed as follows:
\begin{equation}\label{state space trajectory length}
l=\sum_{i=1}^{99}\sqrt{\sum_{j=1}^N \Big[x_j(t_{i})-x_j(t_{i+1}) \Big]^2 },
\end{equation}
where it should be noted that although the time variable $t \in [0,t_f]$ is a continuous variable, it is computationally sampled at $100$ evenly spaced $t_i$ values (for $i=1,2,...,100$). Thus, the length $l$ of the state space trajectory is computationally approximated as the sum of the lengths of $99$ pieces of straight line Euclidean distances between ${\bf x}(t_1)$ and ${\bf x}(t_2)$, ${\bf x}(t_2)$ and ${\bf x}(t_3)$, ..., ${\bf x}(t_{99})$ and ${\bf x}(t_{100})$.

One may wish to ask if a higher $z^*$ is more beneficial or less beneficial to the energy cost needed in controlling the complex network. Plotting $z^*$ against $\mathcal{E}^{\text{large }t_f}_N(z^*)$ in Figs.\ \ref{fig:fig2}(b) and (h), it can be seen that within the same $r$ value, an increase in turning point $z^*$ leads to a decrease in minima energy cost $\mathcal{E}^{\text{large }t_f}_N(z^*)$. Further, increasing $r$ values leads to reduction of minima energy cost, as the more normal agents that are influenced by assisting optimal zealot opinion $z^*$, the less the control signals have to work to drive the network state vector toward consensus. Note that the optimal zealot opinion $z^*$ is not necessarily always the same as consensus $c=5$, and as $r$ value increases, $z^*$ decreases owing to the fact that the network node states follow linear dynamics. In other words, because linear dynamics is always adding or subtracting node states based on the coupled complex networked interactions, having more normal agents influenced by the zealot node leads to the saturation of node states (bias toward the direction of $z$) in the long time. Thus, when $r$ value is relatively large, such as when $r/N=40\%$ or $r/N=60\%$, a zealot opinion of $z=c=5$ will not necessarily assist in lowering energy cost due to the saturation of node states. Instead, the optimal zealot opinion $z^*$ need to be a milder $z$ value below $5$ in order for the zealot's presence to be energy cost-assisting. Finally, (not shown in Fig.\ \ref{fig:fig2}), increasing $r$ value also leads to the $\mathcal{E}^{\text{large }t_f}_N(z)$ quadratic curve having a steeper parabola, and the effective assisting $z$ range where energy cost is lowered (relative to green dashed line) is shortened.

Which network properties of the zealot-influenced nodes control the turning points $z^*$ and associated minima energy costs $\mathcal{E}(z^*)$? To investigate, various common network properties, such as the eigenvalues, degree $\langle k \rangle$, centralities (betweenness, closeness, and eigenvectors), and PageRank \cite{matlab_centrality} of each node are measured and used to distinguish independent configurations of zealot-influenced nodes by selecting them in descending order based on these features. From these measurements, it was found that PageRank and node degree $\langle k \rangle$ of the zealot-influenced nodes are correlated to the turning points $z^*$, where an increase in average PageRank or $\langle k \rangle$ of the zealot-influenced nodes leads to an increase in $z^*$, as shown in Figs.\ \ref{fig:fig2}(c) and (i) (only $\langle k \rangle$ results displayed).

To ensure the sampling of zealot-influenced nodes are well-represented along a diverse $z^*$ range, the configurations of zealot-influenced nodes were chosen according to descending order of nodes degree $\langle k \rangle$. For example, for $r/N=10\%$, the top $10\%$ nodes with highest degree $\langle k \rangle$ were selected, followed by the next $10\%$, and so on. For $r/N=20\%$, $r/N=40\%$, and $r/N=60\%$, the selection is similar, except that between sets, there could be overlapping choices of nodes. Using any other selection strategy other than distinguishing $\langle k \rangle $ (or PageRank) would lead to a tight cluster of data points around similar $z^*$ values in Figs.\ \ref{fig:fig2}(b) and (h). Since node degrees $\langle k \rangle$ correlate to turning points $z^*$, the network topological differences between ER and SF networks in degree distributions $P(k)$ \cite{albert2002statistical} explain the disparity of $z^*$ data points in Figs.\ \ref{fig:fig2}(b) and (h), where the $z^*$ of the SF network tend to be larger, due to the presence of hubs with large average degree $\langle k \rangle$. Physically, this means that when hubs in SF networks have become influenced by the zealot node with assisting $z^*$ opinion, the energy cost is most reduced. Conversely, when hubs are influenced by zealot nodes with $z$ opinion far away from its turning point, the energy cost is most increased as compared to the situation where there are no zealots present.

Figs.\ \ref{fig:fig2}(d)\textendash(f) and (j)\textendash(l) reveal how the zealots sway the state space trajectory lengths $l$ and associated energy costs. For each configuration of zealot-influenced nodes, driving the state vector toward consensus $c=5$, the energy cost and length $l$ are measured respectively for the zealot's $z$ value as $z=-5$ in Figs.\ \ref{fig:fig2}(d) and (j), $z=5$ in Figs.\ \ref{fig:fig2}(e) and (k), and $z=z^*$ in Figs.\ \ref{fig:fig2}(f) and (l). In all of these figures, the computations show that the energy cost is proportional to length $l$: 
\begin{equation}
\mathcal{E}^{\text{large }t_f}_N(l) \propto l.
\end{equation}
With the exception of $z=z^*$ results for $r=1$, $r=2$, $r=3$, and $r=4$ configurations, the data points show that a configuration that leads to a control action that takes a longer $l$ path requires a higher energy cost. Depending on the zealot's $z$ value relative to a specific configuration's $z^*$ value, the zealots' presence may reduce or increase energy cost relative to the energy cost in the absence of zealots. For example, in Figs.\ \ref{fig:fig2}(d) and (j), when $z=-5$, contrarian to the consensus goal, and far away from their respective $r$ values turning point $z^*$, all energy cost increases as compared to the no-zealot energy cost (when $z=0$). Further, at this $z$ value, an increase in $r$ number of zealot-influenced nodes leads to high saturation of zealot-bias node states owing to the networked linear dynamics and the driver nodes would thus need to consume higher energy cost to overcome the zealot's influence. In the converse situation, in Figs.\ \ref{fig:fig2}(e) and (h), when $z=5$, supportive to the consensus goal, most of the energy cost are lowered, compared to the no-zealot energy cost. When $z=5$, for configurations near the associated turning point $z^*$, the zealot's presence is beneficial for control, decreasing energy cost. For the magenta triangles, corresponding to data points of $r/N=60\%$, a $z$ value of $5$ is far away from their associated turning points $z^*$, and in this situation, the zealot's presence is adversarial for control, increasing energy cost. At $z=z^*$, all configurations are at their respective turning points, and the zealots' presence assist in controlling the network, reducing energy cost.


In the small $t_f$ regime, using $N$ drivers, the energy costs (Eqn.\ \ref{E_N small tf}) quadratic curves are plotted in Figs.\ \ref{fig:fig3}(a), (b), (g), and (h) respectively for ER and SF networks with small $t_f=0.1$ and $t_f=0.01$. In these figures, the color-coded plots correspond to the $r$ number of zealot-influenced nodes in accordance to legends in Figs.\ \ref{fig:fig3}(c) or (i). As predicted analytically (Eqn.\ (\ref{z minimum N small tf})), the turning point $z^*_{N \text{, small }t_f} \approx ct_f^{-1}$ is independent of the $r$ number, as well as the choice, of zealot-influenced nodes. From these figures, the theoretical prediction is validated, and all turning points are the same, regardless of $r$ value, and affected only by small $t_f$ value, where $z^*_{N \text{, small }t_f} \approx 50 $ and $z^*_{N \text{, small }t_f} \approx 500 $ respectively for $t_f=0.1$ and $t_f=0.01$, given that $c=5$ is fixed. Its invariant turning point $z^*$ and associated minima energy cost $\mathcal{E}^{\text{small }t_f}_N(z^*)$ are plotted in Figs.\ \ref{fig:fig3}(c) and (i) for varying $r$ values, where an increase in $r$ leads to a decrease in minima energy cost. Unlike its large $t_f$ counterparts, no topological effects between ER and SF networks are observed, as in the small $t_f$ regime, there is barely enough time for the network topological effects such as node degree $\langle k \rangle$ to take effect, and only the $r$ number (and not choice) of influenced nodes  matters to the energy cost.

The energy cost $\mathcal{E}^{\text{small }t_f}_N (l)$ as a function of state space trajectory length $l$ (Eqn.\ (\ref{state space trajectory length})) are computed and plotted in Figs.\ \ref{fig:fig3}(d), (e), (f), (j), (k), and (l). Since the small $t_f$ regime energy cost is invariant to the choice of zealot-influenced nodes, each $r$ value ($r$ denoted by different symbols) only has one data point. To generate more data points to analyze $\mathcal{E}^{\text{small }t_f}_N (l)$ and $l$ dependency, $z$ values were varied ($z$ values denoted by color). Respectively, $z\in [-100,-20]$ for Figs.\ \ref{fig:fig3}(d) and (j), $z\in [20,100]$ for Figs.\ \ref{fig:fig3}(e) and (k), and $z\in [120,200]$ for Figs.\ \ref{fig:fig3}(f) and (l). $z\in [20,100]$ corresponds to the $z$ range where the zealot's presence assists in lowering the energy cost (relative to the no-zealot energy cost), while $z\in [-100,-20]$ and $z\in [120,200]$ are all $z$ values where the zealot's presence would increase energy cost. In Figs.\ \ref{fig:fig3}(d), (f), (j), and (l), state space trajectory $l$ correlates to energy cost, and an increase in $l$ leads to an increase in $\mathcal{E}^{\text{small }t_f}_N(l)$. Further, increasing $r$ value or $|z|$ values also leads to increased $l$ and energy cost, suggesting that when more normal agents are influenced by the zealot node with strong opinions, more effort is needed to steer the network node states toward consensus. In the assisting $z\in [20,100]$ range (Figs.\ \ref{fig:fig3}(e) and (k)), $l$ anti-correlates to  $\mathcal{E}^{\text{small }t_f}_N(l)$, where an increase in $l$ leads to a decrease in energy cost. When $r$ value is the highest, and $z$ value is closest to $z^*=50$, energy cost is most reduced. 


\subsubsection{One driver}

\begin{figure*}
\begin{center} 
\includegraphics[scale=0.64]{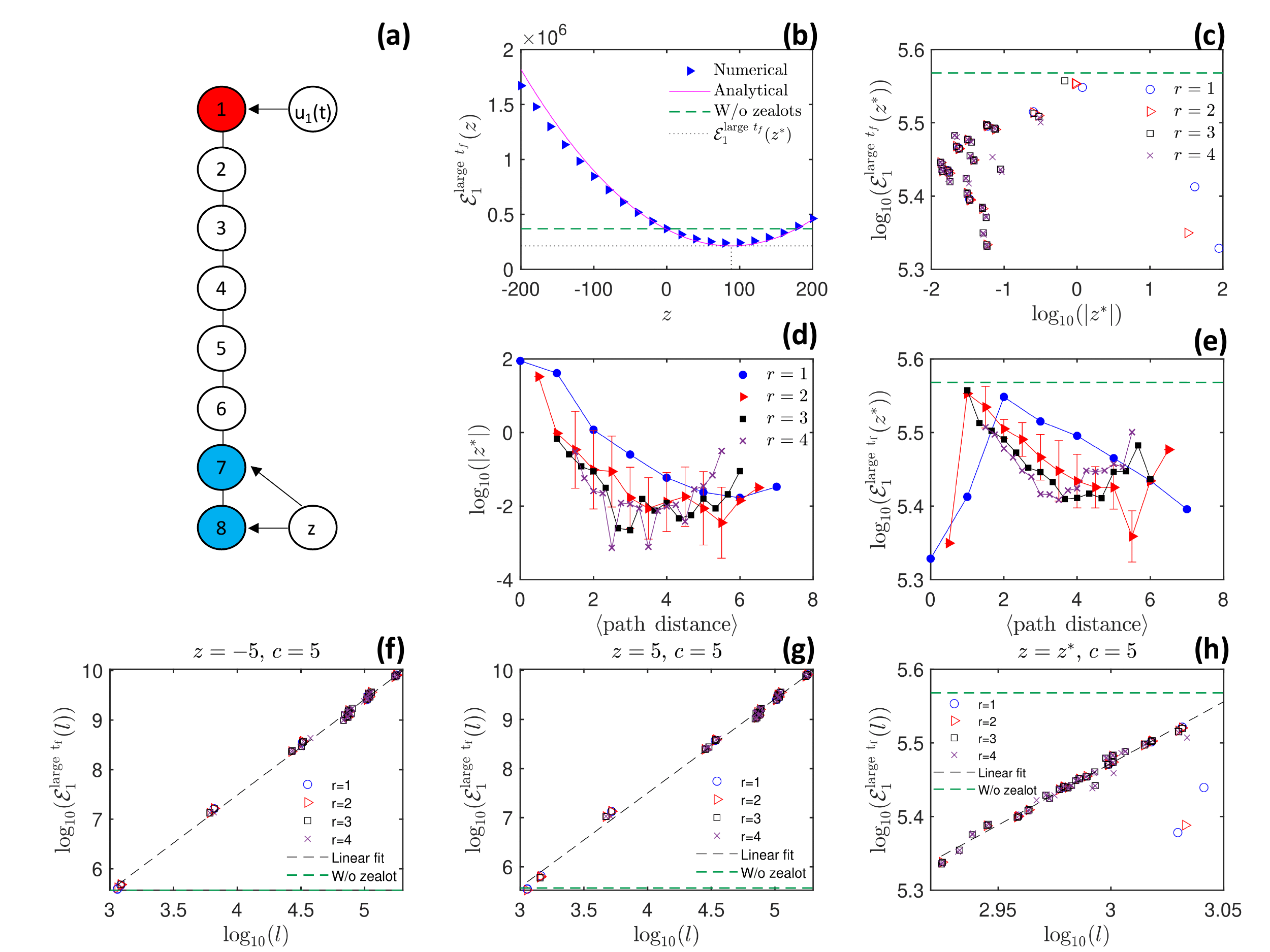} 
\end{center}
\caption{(a) Continuous-time linear dynamics: Large $t_f$ regime, one driver results in a chain network, where the driver node is located at node $1$, and the configurations of zealot-influenced nodes are varied. (b) validates Eqn.\ (\ref{E_1 large tf}). (c) shows how the turning points $\log_{10}(|z^*|)$ relate to its associated minima energy costs $\log_{10}(\mathcal{E}^{\text{large }t_f}_1(z^*))$. Note that unlike the results from $N$ drivers, the turning points $z^*$ could lie on the negative $z$ region, where a contrarian zealot opinion assists the driver node by reducing energy cost. (d) shows that there is a trend where the turning points $z^*$ tend to decrease when the average path distances from driver node $1$ to the zealot-influenced normal agent nodes increase. (e) displays a similar trend, showing that when nodes far away from the driver nodes are influenced by the zealot holding $z^*$ opinion, the energy cost is reduced. Errorbars of $r=2$ (red right triangles)  indicate standard deviations, while the errorbars of $r=3$ (black squares), and $r=4$ (purple crosses) are not shown in plots for the purpose of visibility. (f), (g), and (h) respectively indicate that the energy cost scales with length $l$ according to  $\log_{10}(\mathcal{E}(l))\sim \log_{10}(l)$ for various configurations of zealot-influenced nodes when $z=-5$, $z=5$, and $z=z^*$ when controlling the chain network toward consensus $c=5$. When $z=-5$ or $z=5$, almost all configurations of zealot-influenced nodes increase the energy cost (relative to the energy cost needed for controlling the network in the absence of zealots) because the $z$ values are far from their respective turning points $z^*$. When $z=z^*$, all configurations of zealot-influenced nodes reduce the energy cost} \label{fig:fig4}
\end{figure*}

\begin{figure*} 
\begin{center} 
\includegraphics[scale=0.64]{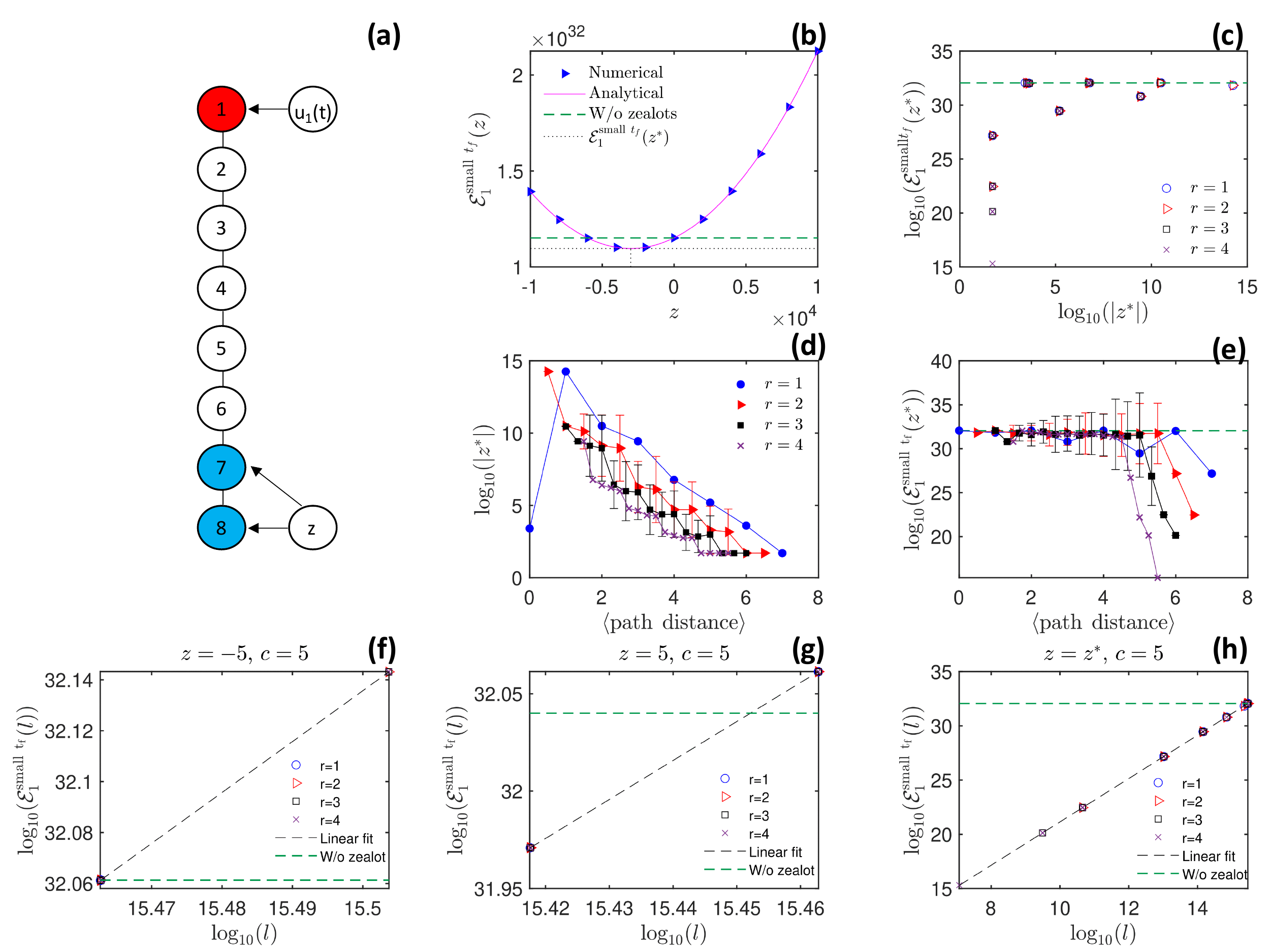} 
\end{center}
\caption{(a) Continuous-time linear dynamics: Small $t_f$ regime, one driver results in a chain network, where the driver node is located at node $1$, and the configurations of zealot-influenced nodes are varied. (b) validates Eqn.\ (\ref{E_1 small t_f and E_d}). (c) shows the relationship between turning points $\log_{10}|z^*|$ and associated minima energy costs $\log_{10}(\mathcal{E}^{\text{small }t_f}_1(z^*))$. (d) shows that as the average path distances from driver node $1$ to the zealot-influenced normal agent nodes increase, the turning points $z^*$ decrease. (e) shows that when normal agents furthest away from the driver node are influenced by the zealot holding $z^*$ opinion, the energy cost is most reduced. (f), (g), and (h) respectively indicate that the energy cost scales with its associated length $l$ as $\log_{10}(\mathcal{E}(l)) \sim \log_{10}(l)$, when $z=-5$, $z=5$, and $z=z^*$ for various configurations of zealot-influenced nodes.  } \label{fig:fig5}
\end{figure*}

When using only one control signal to control the complex network, the various node states are indirectly driven through various paths throughout the network from the sole control signal, and the resulting state space trajectory is highly circuitous \cite{yan2012controlling}. Owing to the highly circuitous state space trajectory, it is difficult to analyze the effects the choice of zealot-influenced nodes have on the energy cost in a complex network. Therefore, in the results that follows, simple network topologies such as chain, ring, and star networks \cite{kafle2018optimal} will be studied instead. 


The one driver large $t_f$ regime results from the numerical experiments with chain, star, and ring networks are presented in Figs.\ \ref{fig:fig4}, and \ref{fig:fig12}\textendash\ref{fig:fig14} (Figs.\ \ref{fig:fig12}\textendash \ref{fig:fig14} in {\bf Appendix \ref{one driver extra results}}). In these figures, the energy costs needed in controlling these networks when there is a zealot node present and influencing $r=1$, $r=2$, $r=3$, and $r=4$ number of normal agents are measured. For each $r$ value, the measurement is repeated for an exhaustive $\Comb{N}{r}$ number of times, each time with different unique sets of zealot-influenced nodes (for small $N$, this search space is feasible). As predicted from Eqn.\ (\ref{E_1 large tf}), the energy cost is quadratic with respect to the zealot's fixed $z$ opinion, which are all validated in Fig.\ \ref{fig:fig4}(b) and Figs.\ \ref{fig:fig12}\textendash \ref{fig:fig14}(b). Unlike its large $t_f$ regime, $N$ drivers counterpart, there is no clear relationship between $z^*$ and $\mathcal{E}(z^*)$, as shown in Figs.\ \ref{fig:fig4}(c), and \ref{fig:fig12}\textendash \ref{fig:fig14}(c). However, in all of these, the statistical trend suggests that when $|z^*|$ is small, minima energy costs $\mathcal{E}^{\text{large } t_f}_1(z^*)$ is most reduced relative to the no-zealot energy cost (horizontal green dashed line). Further, it was noticed during computation that when $|z^*|$ is large, the parabola of the $\mathcal{E}^{\text{large }t_f}_1(z)$ curve stretches, and when $|z^*|$ is small, it steepens. Finally, some of the turning points $z^*$ can lie in the negative $z$ region, indicating that despite holding contrarian opinion to the goal of driving the network toward consensus $c=5$, a contrarian zealot surprisingly assists in the reducing control energy. 

The various network properties of configurations of zealot-influenced nodes were measured to find out if any of them can explain the turning points $z^*$. Unlike its (large $t_f$ regime) $N$ drivers counterparts, node degree $\langle k \rangle$ of zealot-influenced nodes do not predict $z^*$ value for the one driver result. Instead, in Fig.\ \ref{fig:fig4}, for a chain network, where the driver node is located at the root node (node $1$), it was found that the average path distances from driver node to zealot-influenced nodes anti-correlate with $\log_{10}(|z^*|)$, where configurations with zealot-influenced nodes further away from the driver node $1$ tend to have lower $\log_{10}(|z^*|)$, and lower minima energy cost $\log_{10}(\mathcal{E}^{\text{large }t_f}_1(z^*))$. This is not surprising, considering that it has already been reported that for a chain network, energy cost increases exponentially as the path distances increases linearly \cite{chen2016energy}. Thus, when zealots influence nodes that are furthest away from the driver node with assisting $z^*$ opinion, the energy cost can be most reduced.

For more complicated simple network topologies in Figs.\ \ref{fig:fig12}\textendash\ref{fig:fig14}, none of the network properties are predictive of $|z^*|$ nor $\mathcal{E}^{\text{large }t_f}_1(z^*)$. However, in all of these, state space trajectory length $l$ remains a strong predictor of energy cost, yielding the scaling law 
\begin{equation}\label{one driver large tf E(l) vs l scaling law}
\log_{10}(\mathcal{E}^{\text{large }t_f}_1(l)) \sim \log_{10}(l),
\end{equation}
as evidenced in Figs.\ \ref{fig:fig4}(f)\textendash(h), and Figs.\ \ref{fig:fig13}\textendash \ref{fig:fig15}(d)\textendash (f), respectively for $z=-5$, $z=5$, and $z=z^*$. Thus, when a particular configuration of zealot-influenced nodes leads to low length $l$, energy cost is most reduced. When $z=-5$ or $z=5$, most of the configurations result in increased energy cost relative to no-zealot energy cost (green dashed lines) as the $z$ values are far from their turning points $z^*$. When $z=z^*$, all configurations lead to reduced energy cost. In star or ring network, the most optimal configurations can reduce the energy cost by $3$-$4$ orders of magnitude. Finally, note that most of the scatter data points overlap, and there is no clear distinction between $r$ values, indicating that for one driver, large $t_f$ regime, the zealot's influencing of $1$ normal agent has as much potential in swaying the control action as influencing $4$ normal agents. This suggests that, because the one driver state space trajectory is highly circuitous \cite{yan2012controlling}, $r$ number of nodes being influenced by the zealot is not predictive of how much the energy cost will change since the circulation of the zealot's influence may move the state space trajectory around through different indirect ways.


The numerical experiments are repeated for the one driver node calculations in the small $t_f$ regime. Respectively, Fig.\ \ref{fig:fig5}, and Figs.\ \ref{fig:fig16}\textendash\ref{fig:fig18} (Figs.\ \ref{fig:fig16}\textendash\ref{fig:fig18} are found in {\bf Appendix \ref{one driver extra results}}) correspond to chain network with node $1$ being the driver node, chain network with node $4$ being the driver node, star network, and ring network. Similar to its large $t_f$ counterparts, there is no clear trend between turning points $\log_{10}(|z^*|)$ and minima $\log_{10}(\mathcal{E}^{\text{small }t_f}_1(z^*))$, except that it appears that a smaller $|z^*|$ value tend to have a lower $\mathcal{E}^{\text{small }t_f}_1(z^*)$. Further, the turning points $z^*$ can also lie in the negative $z$ region, indicating that a contrarian zealot can actually assist in lowering the energy cost. Overall, the physical behavior of the one driver node computations in the small $t_f$ regime is similar to its large $t_f$ regime counterparts, except that the required energy is much larger, which is not surprising, given that the required energy cost decreases as $t_f$ increases \cite{yan2012controlling}. The scaling behavior 
\begin{equation}\label{one driver small tf E(l) vs l scaling law}
\log_{10}(\mathcal{E}^{\text{small }t_f}_1(l)) \sim \log_{10}(l)
\end{equation}
are corroborated by Fig.\ \ref{fig:fig5}(f)\textendash(h), and Figs.\ \ref{fig:fig16}\textendash \ref{fig:fig18}(d)\textendash(f).


\subsubsection{d drivers}

\begin{figure*} 
\begin{center} 
\includegraphics[scale=0.41]{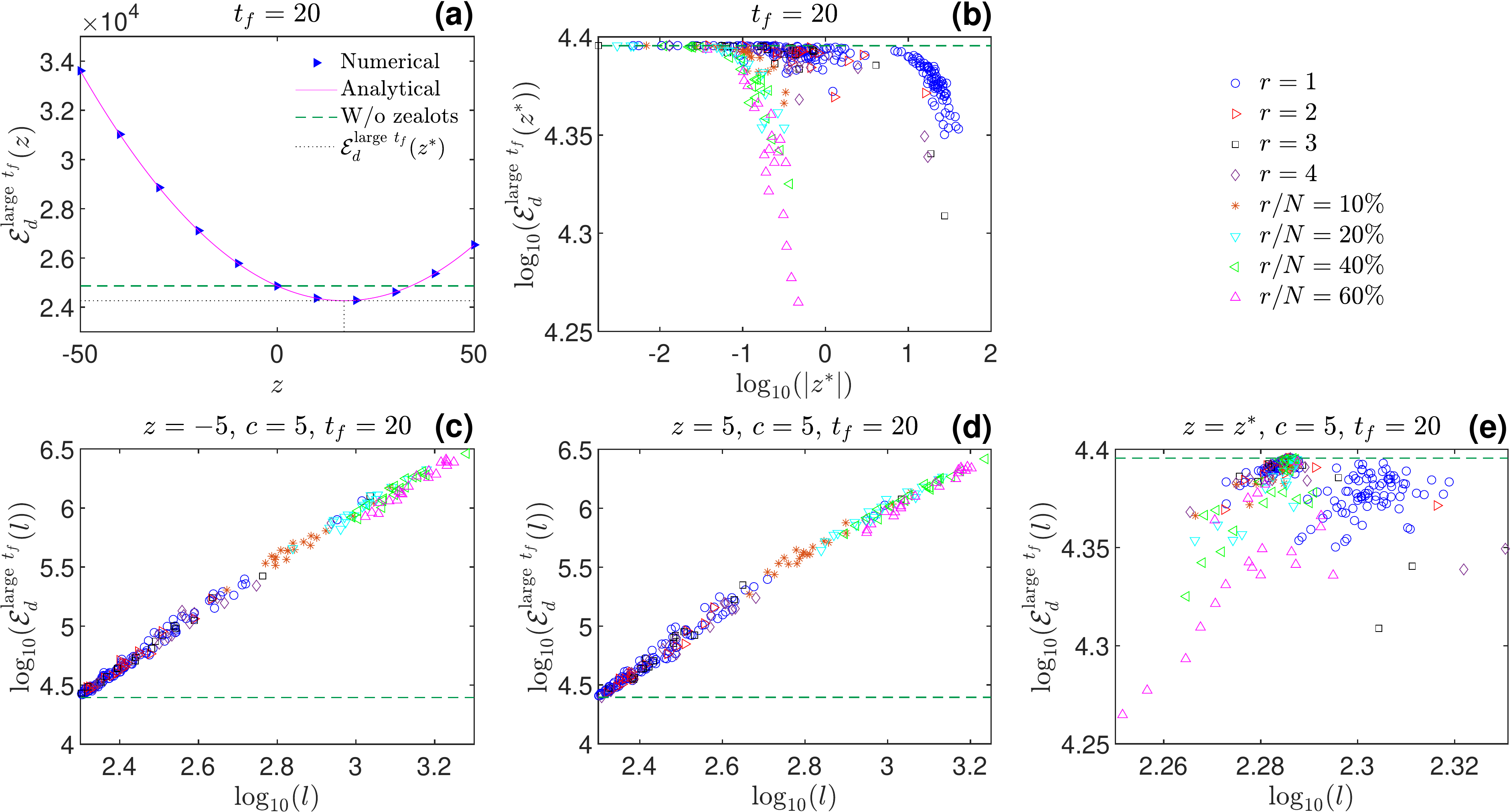} 
\end{center}
\caption{Continuous-time linear dynamics: Large $t_f$ $d$ drivers results in a $N=200$, $\langle k \rangle=6$ ER network with $80$ control signals. (a) validates Eqn.\ (\ref{E_1 small t_f and E_d}), showing that the energy cost is quadratic with respect to zealot's $z$ opinion. (b) shows the relationship between turnings $z^*$ and their associated minima energy cost $\mathcal{E}_d(z^*)$. (c)\textendash(e) plot the state space trajectory length $l$ that each zealot-influenced nodes configuration takes the system and the corresponding energy cost at $z=-5$, $z=5$, and at $z=z^*$. They show that the energy cost of each configuration can be explained by length $l$, where the energy cost scales as $\log_{10}(\mathcal{E}_d(l))\sim \log_{10}(l)$. Legend displays the marker symbols corresponding to each $r$ number of zealot-influenced normal agents.  } \label{fig:fig6}
\end{figure*}

\begin{figure*} 
\begin{center} 
\includegraphics[scale=0.41]{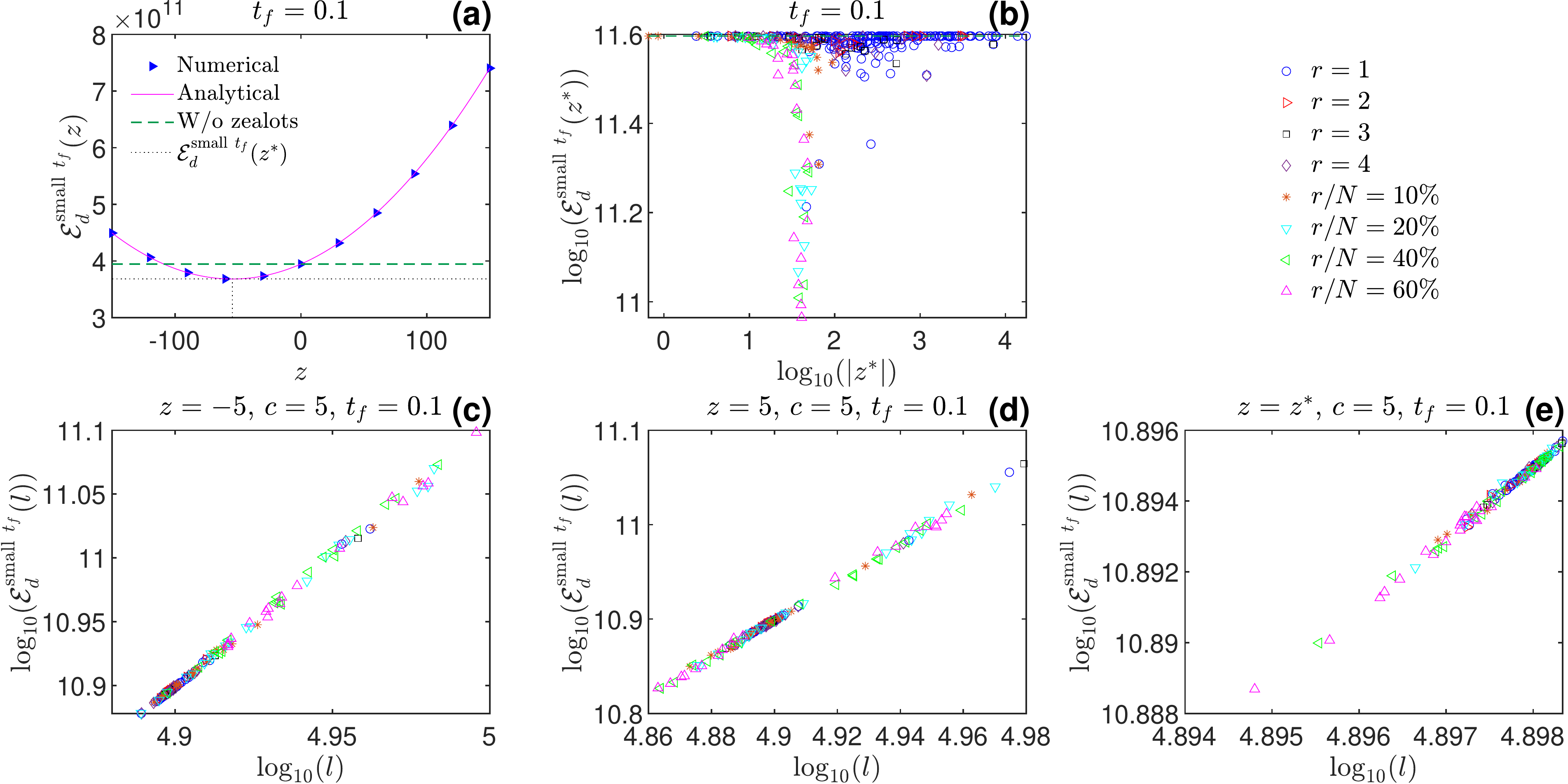} 
\end{center}
\caption{Continuous-time linear dynamics: Small $t_f$ $d$ drivers results in a $N=200$, $\langle k \rangle=6$ ER network with $80$ control signals. (a) validates Eqn.\ (\ref{E_1 small t_f and E_d}), showing that the energy cost is quadratic with respect to zealot's $z$ opinion. (b) shows the relationship between turnings $z^*$ and their associated minima energy cost $\mathcal{E}_d(z^*)$. (c)\textendash(e) plot the state space trajectory length $l$ that each zealot-influenced nodes configuration takes the system and the corresponding energy cost at $z=-5$, $z=5$, and at $z=z^*$. They show that the energy cost of each configuration can be explained by length $l$, where the energy cost scales as $\log_{10}(\mathcal{E}_d(l))\sim \log_{10}(l)$. In the small $t_f$ regime, all configurations reduce the energy cost relative to no-zealot energy cost (not shown in Figs.\ (c)\textendash(e)), which was $\log_{10}(\mathcal{E})\approx 11.6$. Legend displays the marker symbols corresponding to each $r$ number of zealot-influenced normal agents. } \label{fig:fig7}
\end{figure*}

The numerical experiments were repeated for controlling a $N=200$, $\langle k \rangle = 6$ ER network (it is expected that similar results hold for SF networks) with continuous-time linear dynamics using $80$ control signals. The results, both large $t_f$ and small $t_f$ regimes are plotted respectively in Figs.\ \ref{fig:fig6} and \ref{fig:fig7}. Figs.\ \ref{fig:fig6}(a) and \ref{fig:fig7}(a) validate Eqn.\ (\ref{E_1 small t_f and E_d}), showing that the energy cost is quadratic with respect to the zealot's fixed $z$ opinion. Furthermore, Fig.\ \ref{fig:fig7}(a) shows that the turning point $z^*$ lie in the negative $z$ region, although the goal is to drive the state vector towards $c=5$, indicating that a contrarian zealot is beneficial for control in some zealot-influenced nodes configurations. There is no clear relationship between turning points $z^*$ and their associated minima energy costs $\mathcal{E}(z^*)$, as displayed in Figs.\ \ref{fig:fig6}(b) and \ref{fig:fig7}(b). This is in contrast to the results of large $t_f$ $N$ drivers, where an increase in $z^*$ leads to a decrease in $\mathcal{E}(z^*)$, and the one driver results, where a decrease in $z^*$ tend to decrease $\mathcal{E}(z^*)$. 

Nevertheless, the energy cost of each zealot-influenced nodes configuration can be explained by the state space trajectory length $l$. $l$ was calculated, along with its associated energy cost, for fixed $z=-5$, $z=5$, and $z=z^*$, where for each $r$ number of zealot-influenced nodes, from $r=2$ to $r/N=60\%$, $20$ independent configurations of nodes were randomly selected. For $r=1$, all independent configurations from node $1$ to node $N$ were chosen. From Figs.\ \ref{fig:fig6}(c)\textendash(e) and \ref{fig:fig7}(c)\textendash(e), it is evident that the energy cost scales as
\begin{equation}
\log_{10}(\mathcal{E}_d(l)) \sim \log_{10}(l),
\end{equation}
where a configuration that causes an increased $l$ would cost the drivers more energy cost. The energy that each configuration costs is relative to their respective turning point $z^*$ and the steepness of the parabola. For the large $t_f$ results, $z=-5$ and $z=5$ configurations are far from the minima and all configurations have higher energy costs compared to the no-zealot energy cost (green dashed line). At $z=z^*$, all configurations are exactly at their minima, and all configurations reduce energy cost, although by less than one order of magnitude. In the small $t_f$ regime, for $z=-5$, $z=5$, and $z=z^*$, all configurations are close to or at the minima, and the zealot's influence reduces the energy cost as compared to no-zealot energy cost (not shown in Figs.\ \ref{fig:fig7}(c)\textendash(e)), which was around $\log_{10}(\mathcal{E})\approx 11.6$. 

\section{Discrete-time linear dynamics with conformity behavior model} \label{discrete-time model with conformity subsection}
Next, how zealots affect the energy cost in controlling a complex network with discrete-time linear dynamics and conformity behavior \cite{wang2015controlling} is studied. The dynamics of this model incorporates conformity features, where each member of the network adapts their node state to follow the average of their nearest neighbors over time. Arguably, it is more realistic as it captures conformity behavior, which has been reported in various social systems. For example, evolutionary games \cite{wu2014role,hilbe2014cooperation}, learning behaviors \cite{van2013potent,whiten2005conformity}, and collective movements \cite{vicsek2012collective,vicsek1995novel,nagy2010hierarchical,ward2008quorum,buhl2006disorder} of animals in groups. 

\begin{figure*} 
\begin{center} 
\includegraphics[scale=0.64]{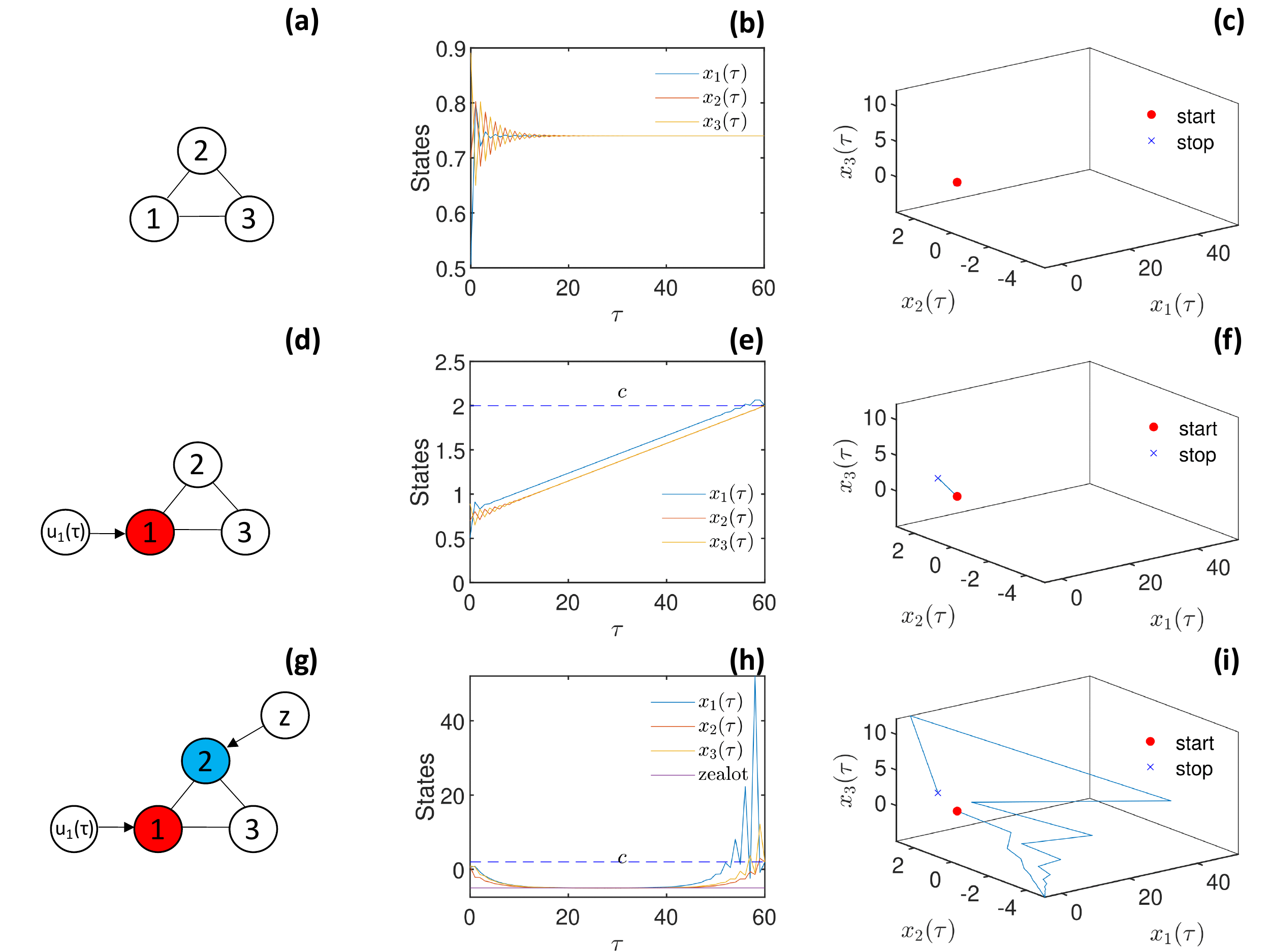}
\end{center}
\caption{(a) Network with conformity behavior in the absence of control signal and zealot. (b) Network node states evolution with conformity in the absence of external influences. (c) State space trajectory of $x_1(\tau)$, $x_2(\tau)$, and $x_3(\tau)$ in the absence of external influences. (d) Network with conformity behavior and control signal $u_1(t)$ attached to node $1$. (e) Conformity-based network node states being driven toward consensus $c=2$. (f) State space trajectory of nodes being driven toward consensus. (g) Network with conformity behavior, control signal $u_1(t)$ attached to node $1$, and zealot node influencing node $2$. (h) Conformity-based network node states being driven toward consensus, while node $2$ is under the influence of the zealot node with fixed opinion $z=-5$. (i) State space trajectory of nodes being driven toward consensus while under the influence of the zealot.} \label{fig:fig8}
\end{figure*}

An example is given in Fig.\ \ref{fig:fig8} to introduce this network dynamics, where the node states $x_i(t)$ may represent different opinions on a particular subject, with a positive $x_i(t)$ value indicating support of an idea, and a negative $x_i(t)$ value representing opposition. A minimum of three nodes are needed to showcase the conformity dynamics because otherwise, with two nodes, they will just mimic each other's node states in perpetuity without reaching conformity. In Fig.\ \ref{fig:fig8}(a), there are $N=3$ normal agents who start with different node states $x_i(\tau=0)$ drawn from random uniform $[0,1]$. Over time, each normal agent mimics their nearest neighbors' node states and the network node states reach conformity as demonstrated in Fig.\ \ref{fig:fig8}(b). The $3$-dimensional state space trajectory of this time evolution of node states is presented in Fig.\ \ref{fig:fig8}(c). In Fig.\ \ref{fig:fig8}(d), a single control signal $u_1(\tau)$ attaches to node $1$ to control the network, driving the node states toward consensus $c=2$, as shown in Fig.\ \ref{fig:fig8}(e). The state space trajectory of this control action is plotted in Fig.\ \ref{fig:fig8}(f). Finally, in Fig.\ \ref{fig:fig8}(g), a zealot node is introduced into the system, where it holds a fixed $z=-5$ opinion. Fig.\ \ref{fig:fig8}(h) shows the node states evolution of the normal agents being driven toward $c=2$, along with the zealot's fixed opinion at $z=-5$. Correspondingly, in Fig.\ \ref{fig:fig8}(i), the state space trajectory of this control action, under the influence of the zealot, elongates as the control signal $u_1(\tau)$ now has to consume more energy to overcome the zealot's contrarian opinion. Note that it suffices to display the state space trajectory in $3$ dimensions as the zealot's fixed node state, associated with the fourth dimension, would simply shift and constraint $x_1(\tau)$, $x_2(\tau)$, and $x_3(\tau)$ onto the fixed $x_4=z$ hyperplane.

Networked conformity dynamics is achieved when, at the individual level, each normal agent node updates their node states in the next discrete-time round, $\tau +1$, by taking the average of their nearest neighbors' in the current round $\tau$ \cite{wang2015controlling}. Therefore, for each normal agent node $i$ (for $i=1,2,...,N$),
\begin{equation}\label{individual level conformity}
x_i(\tau+1)=\frac{1}{\tilde{s}_i} \sum_{j=1}^{\tilde{n}_i} \tilde{a}_{ij} x_j(\tau),
\end{equation}
where node $i$ has $\tilde{n}_i$ nearest normal agent neighbors, $x_j(\tau)$ is the opinion of neighbor $j$ at round $\tau$, $\tilde{a}_{ij}=\tilde{a}_{ji}$ is the weighted connection between nodes $i$ and $j$, and $\tilde{s}_i=\sum\limits_{j=1}^{\tilde{n}_i} \tilde{a}_{ij}$ is the strength of node $i$, obtained by summing over all nearest neighbors' weighted connections. Introducing input control signal terms into Eqn.\ (\ref{individual level conformity}), the zealot's connections, and rewriting in vector notational form, the system-level dynamics is
\begin{equation} \label{discrete-time LTI conformity dynamics}
\begin{aligned}
{\bf x}(\tau +1) = & {\bf S}^{-1}{\bf A} {\bf x}(\tau) + {\bf B}{\bf u}(\tau)\\
=& \ubar{\bf A}{\bf x}(\tau) + {\bf B}{\bf u}(\tau),
\end{aligned}
\end{equation}
where ${\bf x}(\tau)=[x_1(\tau),x_2(\tau),...,x_N(\tau),z]^T$ is the $n \times 1$ state vector of all $N$ normal agents, and the zealot node (the $n$-th node, where $n=N+1$), ${\bf u}(\tau)=[u_1(\tau), u_2(\tau), ..., u_M(\tau)]^T$ is the $M \times 1$ control signals vector, ${\bf A}$ is the full $n \times n$ network matrix, which comprises $\tilde{\bf A}$ as its first $N \times N$ block, where $\tilde{a}_{ij}=\tilde{a}_{ji}$ is non-zero if normal agents $i$ and $j$ have interactions, otherwise it is zero, and the $n$-th column describes the zealot's directed link to normal agents, where $a_{in}=1$ if zealot node influences node $i$, otherwise it is zero, ${\bf B}$ is the control input matrix which describes which nodes are directly controlled by a control signal such that $b_{ij}=1$ if normal agent node $i$ is controlled by control signal $j$, otherwise it is zero, ${\bf S}^{-1}=\text{diag}\{\frac{1}{s_1},\frac{1}{s_2},...,\frac{1}{s_N},1\}$ is a $n \times n$ diagonal matrix which holds $\frac{1}{s_i}$ (for $i=1,2,...,N$) on its main diagonals, with the $n$-th entry being $1$, and $s_i = \sum\limits_{j=1}^{n_i}a_{ij}$ sums over the full ${\bf A}$ matrix, inclusive of the zealot node's directed connections. $\ubar{\bf A}={\bf S}^{-1} {\bf A}$ is the full network matrix, where the normal agents evolve in time with conformity, yet the zealot node remains fixed with state $z$. Note that unlike the continuous-time dynamics, which requires self-links to model system stability, the discrete-time dynamics with conformity model requires that $\tilde{a}_{ii}=0$ to ensure system stability (see {\bf Appendix \ref{appendix methods}}). Further, the zealot's self-loop $[{\bf S}^{-1}]_{nn}=a_{nn}=\ubar{a}_{nn}=1$ is necessary for modelling zealotry, leading to $x_n(\tau+1)=x_n(\tau)=z$, and the zealot node's opinion remains fixed at $z$ against time. Finally, the non-symmetric full network matrix can be eigen-decomposed as $\ubar{\bf A}=\ubar{\bf P} \ubar{\bf D} \ubar{\bf P}^{-1}$ and $\ubar{\bf A}^T = \ubar{\bf V}\ubar{\bf D} \ubar{\bf V}^{-1}$, where $\ubar{\bf P}$ ($\ubar{\bf V}$) is the eigenvectors matrices of $\ubar{\bf A}$ ($\ubar{\bf A}^T$), and $\ubar{\bf D}$ is the $n\times n$ diagonal matrix containing the eigenvalues of $\ubar{\bf A}$ such that $\ubar{\bf D}=\text{diag}\{\Lambda_1,\Lambda_2, ..., \Lambda_N,\Lambda_n \}$, where $\Lambda_1 \leq \Lambda_2 \leq ... \leq \Lambda_N \leq \Lambda_n=1$. From computation, regardless of network size or topology, the eigenvalues of all normal agents $|\Lambda_i|<1$, while the zealot node has eigenvalue $\Lambda_n=1$.

\subsection{Analytical equations of energy cost}

The energy cost required to control the network with discrete-time linear dynamics and conformity behavior is \cite{duan2019target,lewis2012optimal,li2017control,li2017fundamental} 
\begin{equation} \label{discrete-time energy cost function}
J=\sum\limits_{\tau=0}^{T_f-1}u^T(\tau)u(\tau),
\end{equation}
where $T_f$ is the final control time, which is the amount of time allocated to the control signals to steer the state vector. Minimizing the cost function (Eqn.\ (\ref{discrete-time energy cost function})), the discrete-time energy-optimal control signal that should be used to control the network is derived \cite{duan2019target,lewis2012optimal} as the $M \times 1$ vector
\begin{equation} \label{discrete-time energy-optimal control signal}
{\bf u}^*(\tau)={\bf B}^T(\ubar{\bf A}^T)^{T_f-\tau-1}{\bf C}^T ({\bf C}\ubar{\bf W}{\bf C}^T)^{-1}({\bf y}_f - {\bf C}\ubar{\bf A}^{T_f}{\bf x}_0),
\end{equation}
where $\ubar{\bf W}=\sum\limits_{\tau=0}^{T_f-1}\ubar{\bf A}^{T_f-\tau-1}{\bf B}{\bf B}^T (\ubar{\bf A}^T)^{T_f-\tau-1} $ is the $n\times n$ discrete-time controllability Gramian of the conformity-based system, ${\bf x}_0=[x_1(0), x_2(0), ..., x_N(0),z]^T=[0,0,...,0,z]^T$ is the $n\times 1$ initial state vector, where the initial node states of the normal agents are assumed to all be zeros, ${\bf y}_f=[c,c,...,c]^T$ is the $N\times 1$ final output state vector, where normal agents are all controlled toward consensus $c$, and ${\bf C}$ is the $N\times n$ target control matrix which selects the output state vector to be controlled, which in this case are all the $N$ normal agents, so ${\bf C}_{ii}=1$ (for $i=1,2,...,N$), and the $n$-th final row/column of ${\bf C}$ are all zeros.

Thus, substituting Eqn.\ (\ref{discrete-time energy-optimal control signal}) into Eqn.\ (\ref{discrete-time energy cost function}), the required energy cost when using ${\bf u}^*(\tau)$ to control the network is
\begin{equation} \label{discrete-time energy cost generic}
\begin{aligned}
\mathcal{E}=&{\bf y}_f^T({\bf C}\ubar{\bf W}{\bf C}^T)^{-1}{\bf y}_f - 2{\bf x}_0^T(\ubar{\bf A}^T)^{T_f}{\bf C}^T({\bf C}\ubar{\bf W}{\bf C}^T)^{-1}{\bf y}_f\\
& + {\bf x}_0^T (\ubar{\bf A}^T)^{T_f} {\bf C}^T ({\bf C}\ubar{\bf W}{\bf C}^T)^{-1}{\bf C}\ubar{\bf A}^{T_f}{\bf x}_0.
\end{aligned}
\end{equation}
This can be further simplified by noting that \cite{chen2021energy}
\begin{equation}
({\bf C}\ubar{\bf W}{\bf C}^T)= \ubar{\tilde{\bf W}}=\ubar{\tilde{\bf P}}\tilde{\ubar{\bf M}}\tilde{\ubar{\bf V}}^{-1}= \ubar{\tilde{\bf P}}[\tilde{\ubar{\bf Q}}\circ\tilde{\ubar{\bf F}}]\tilde{\ubar{\bf V}}^{-1},
\end{equation}
where the ${\bf C}$ and ${\bf C}^T$ matrices remove the $n$-th row/column of $\ubar{\bf W}$, leading to $\tilde{\ubar{\bf W}}= \sum\limits_{\tau=0}^{T_f-1}\tilde{\ubar{\bf A}}^{T_f-\tau-1} \tilde{\bf B} \tilde{\bf B}^T (\tilde{\ubar{\bf A}}^T)^{T_f-\tau-1}$, the $N\times N$ reduced controllability Gramian matrix. $\tilde{\ubar{\bf A}}$ is the reduced network matrix, obtained by removing the $n$-th row and column off $\ubar{\bf A}$, so its eigen-decompositions are $\tilde{\ubar{\bf A}}=\tilde{\ubar{\bf P}} \tilde{\ubar{\bf D}} \tilde{\ubar{\bf P}}^{-1}$ and $\tilde{\ubar{\bf A}}^T=\tilde{\ubar{\bf V}} \tilde{\ubar{\bf D}} \tilde{\ubar{\bf V}}^{-1}$, where $\tilde{\ubar{\bf P}}$ ($\tilde{\ubar{\bf V}}$) is the eigenvectors matrix of the reduced matrix $\ubar{\tilde{\bf A}}$ ($\tilde{\ubar{\bf A}}^T$), with $\tilde{\ubar{\bf D}}_{ii}=\Lambda_i$ corresponding to the eigenvalue of the normal agent node. Unlike its continuous-time counterpart, the reduced network matrix of the discrete-time dynamical system cannot decouple the zealots' connections owing to ${\bf S}^{-1}{\bf A}$ in Eqn.\ (\ref{discrete-time LTI conformity dynamics}), and so it must be emphasized that $\ubar{\tilde{\bf A}}$ is not the same as $\tilde{\bf A}$. $\tilde{\ubar{\bf M}}$ is the simplified controllability Gramian matrix of the reduced discrete-time system, which has the Hadamard product analytical form \cite{yan2012controlling,yan2015spectrum,duan2019energy,duan2019target,chen2021energy} 
\begin{equation} \label{M_ij conformity}
\ubar{\tilde{\bf M}}_{ij} = \ubar{\tilde{\bf Q}}_{ij} \ubar{\tilde{\bf F}}_{ij}= [\ubar{\tilde{\bf P}}^{-1}\tilde{\bf B}\tilde{\bf B}^T \tilde{\ubar{\bf V}}]_{ij} \Bigg[\frac{1-(\Lambda_i \Lambda_j)^{T_f}}{1-\Lambda_i \Lambda_j} \Bigg].
\end{equation}
Applying the inverse operation, 
\begin{equation} \label{CWCT inverse}
({\bf C}\ubar{\bf W}{\bf C}^T)^{-1} = \tilde{\ubar{\bf W}}^{-1}=(\ubar{\tilde{\bf P}}\tilde{\ubar{\bf M}}\tilde{\ubar{\bf V}}^{-1})^{-1}= \ubar{\tilde{\bf V}}\tilde{\ubar{\bf M}}^{-1}\tilde{\ubar{\bf P}}^{-1}.
\end{equation}


\subsubsection{Large $T_f$ regime}

The scaling laws of the energy cost needed to control networks with conformity behavior with respect to $T_f$ have been studied in Ref.\ \cite{chen2021energy}. Like the continuous-time system \cite{yan2012controlling,duan2019energy}, the upper bound of the energy cost is characterized into two distinct regimes: The small $T_f$ regime and the large $T_f$ regime.  Substituting the eigen-decompositions $\ubar{\bf A}=\ubar{\bf P}\ubar{\bf D} \ubar{\bf P}^{-1} $, $\ubar{\bf A}^T=\ubar{\bf V}\ubar{\bf D} \ubar{\bf V}^{-1}$, and Eqn. (\ref{CWCT inverse}) into Eqn.\ (\ref{discrete-time energy cost generic}), the large $T_f$ energy cost is
\begin{equation} \label{large Tf discrete-time energy cost generic quadratic z} 
\begin{aligned}
\mathcal{E}^{\text{large }T_f}=& c^2 \sum_{i=1}^{N} \sum_{j=1}^{N} [\tilde{\ubar{\bf V}} \tilde{\ubar{\bf M}}^{-1} \tilde{\ubar{\bf P}}^{-1} ]_{ij} \\
& -2cz \sum_{i=1}^{N} \sum_{j=1}^{N} [\tilde{\ubar{\bf V}} \tilde{\ubar{\bf M}}^{-1} \tilde{\ubar{\bf P}}^{-1} ]_{ij} \\
& + z^2 \sum_{j=1}^{N} [\tilde{\ubar{\bf V}} \tilde{\ubar{\bf M}}^{-1} \tilde{\ubar{\bf P}}^{-1} ]_{ij}\\
=& \sum_{i=1}^{N} \sum_{j=1}^{N} [\tilde{\ubar{\bf V}} \tilde{\ubar{\bf M}}^{-1} \tilde{\ubar{\bf P}}^{-1} ]_{ij} \big(c-z \big)^2,
\end{aligned}
\end{equation}
which is quadratic with respect to the zealot's fixed $z$ opinion. The zealot's connections are encoded in $[\tilde{\ubar{\bf V}} \tilde{\ubar{\bf M}}^{-1} \tilde{\ubar{\bf P}}^{-1}]_{ij}$, and the effects of varying the driver nodes connections are encoded in $\tilde{\ubar{\bf M}}^{-1}$. Note that a minimum of one connection from zealot to normal node must be made in order for Eqn.\ (\ref{large Tf discrete-time energy cost generic quadratic z}) to be valid. Since the normal agents have eigenvalues $|\Lambda_i|<1$ (for $i=1,2,...,N$), all $(\Lambda_i\Lambda_j)^{T_f}$ terms in Eqn. (\ref{M_ij conformity}) vanish in the large $T_f$ limit, and the simplified controllability Gramian becomes
\begin{equation}\label{M_ij conformity large Tf}
\ubar{\tilde{\bf M}}_{ij} =  \frac{[\ubar{\tilde{\bf P}}^{-1}\tilde{\bf B}\tilde{\bf B}^T \tilde{\ubar{\bf V}}]_{ij}}{1-\Lambda_i \Lambda_j}.
\end{equation}
Subsequently, the one driver $\tilde{\ubar{\bf M}}^{-1}$ follows similarly from Ref.\ \cite{chen2021energy}, and the one driver energy cost is fully analytical by substituting Eqn.\ (\ref{one driver M inverse}) into Eqn.\ (\ref{large Tf discrete-time energy cost generic quadratic z}). When using more than one driver node, $\tilde{\ubar{\bf M}}^{-1}$ cannot be derived analytically \cite{chen2021energy}, and is computed numerically through the inverse of Eqn. (\ref{M_ij conformity large Tf}).

Taking the derivative $\frac{\partial \mathcal{E}}{\partial z}=0$, the optimal zealot opinion that assists in lowering the energy cost is
\begin{equation}\label{large Tf conformity z turning point}
z^*_{\text{large }T_f} = c,
\end{equation}
regardless of number of drivers. This is unsurprisingly, since the networked conformity dynamics (see Fig.\ \ref{fig:fig8}(b)) is global, and the unwavering $z$ opinion forces the adaptive normal agents to take its value. When $z^*_{\text{large }T_f}=c$, the zealot forces the normal agents to have the consensus $c$ opinion, thereby assisting the driver nodes in their tasks in controlling the network. Substituting $z=c$ into Eqn.\ (\ref{large Tf discrete-time energy cost generic quadratic z}), it is easy to verify that, in fact, the energy cost $\mathcal{E}^{\text{large} T_f}(z^*)$ is zero. 


\subsubsection{Small $T_f$ regime}

In the small $T_f$ regime, the normal agent nodes have less time to adapt to the zealot's $z$ opinion. Substituting the eigen-decompositions $\ubar{\bf A}=\ubar{\bf P}\ubar{\bf D} \ubar{\bf P}^{-1} $, $\ubar{\bf A}^T=\ubar{\bf V}\ubar{\bf D} \ubar{\bf V}^{-1}$, and Eqn. (\ref{CWCT inverse}) into Eqn.\ (\ref{discrete-time energy cost generic}),
\begin{equation} \label{small Tf discrete-time energy cost generic quadratic z} 
\begin{aligned}
\mathcal{E}^{\text{small }T_f}= & c^2 \sum_{i,j}  [\tilde{\ubar{\bf V}} \tilde{\ubar{\bf M}}^{-1} \tilde{\ubar{\bf P}}^{-1}]_{ij} - 2cz \sum_{i,j} [(\ubar{\bf A}^T)^{T_f}]_{ni}[\tilde{\ubar{\bf V}} \tilde{\ubar{\bf M}}^{-1} \tilde{\ubar{\bf P}}^{-1}]_{ij}\\
& + z^2 \sum_{i,j} [(\ubar{\bf A}^T)^{T_f}]_{ni}[\tilde{\ubar{\bf V}} \tilde{\ubar{\bf M}}^{-1} \tilde{\ubar{\bf P}}^{-1}]_{ij} [(\ubar{\bf A})^{T_f}]_{jn}\\
=& c^2 \sum_{i,j}  [\tilde{\ubar{\bf V}} \tilde{\ubar{\bf M}}^{-1} \tilde{\ubar{\bf P}}^{-1}]_{ij}\\
& - 2cz \sum_{i,j,k}\ubar{v}_{ni}[\ubar{\bf V}^{-1}]_{ij}[\tilde{\ubar{\bf V}} \tilde{\ubar{\bf M}}^{-1} \tilde{\ubar{\bf P}}^{-1}]_{jk}\Lambda_i^{T_f}\\
& + z^2 \sum_{i,j,k,l} \ubar{v}_{ni} [\ubar{\bf V}^{-1}]_{ij}\ubar{p}_{kl} [\ubar{\bf P}^{-1}]_{ln}[\tilde{\ubar{\bf V}} \tilde{\ubar{\bf M}}^{-1} \tilde{\ubar{\bf P}}^{-1}]_{jk} \Lambda_i^{T_f} \Lambda_l^{T_f},
\end{aligned}
\end{equation}
where $\sum\limits_{i,j}=\sum\limits_{i=1}^{N}\sum\limits_{j=1}^{N}$, $\sum\limits_{i,j,k}=\sum\limits_{i=1}^{n}\sum\limits_{j=1}^{N}\sum\limits_{k=1}^{N}$, $\sum\limits_{i,j,k,l}=\sum\limits_{i=1}^{n}\sum\limits_{j=1}^{N}\sum\limits_{k=1}^{N}\sum\limits_{l=1}^{n}$, with $i$, $j$, $k$, $l$ as the running indices that iterate from $1$ to $N$ or $1$ to $n$, and $n$ is the fixed $n$-th index. The choice of zealot-influenced nodes enters the energy cost equation through the $n\times n$ matrices $\ubar{v}_{ni}$, $[\ubar{\bf V}^{-1}]_{ij}$, $\ubar{p}_{kl}$, $[\ubar{\bf P}^{-1}]_{ln}$, and the $N \times N$ matrices $\tilde{\ubar{\bf V}} \tilde{\ubar{\bf M}}^{-1} \tilde{\ubar{\bf P}}^{-1}$.

Thereafter, the derivative $\frac{\partial \mathcal{E}}{\partial z}=0$ yields the turning point
\begin{equation}\label{small Tf discrete-time turning point z*}
z^{*}_{\text{small }T_f}=\frac{c \sum\limits_{i,j,k} \ubar{v}_{ni} [\ubar{\bf V}^{-1}]_{ij} [\ubar{\tilde{\bf V}} \ubar{\tilde{\bf M}}^{-1} \ubar{\tilde{\bf P}}^{-1} ]_{jk} \Lambda_{i}^{T_f} }{\sum\limits_{i,j,k,l} \ubar{v}_{ni} [\ubar{\bf V}^{-1}]_{ij} \ubar{p}_{kl} [\ubar{\bf P}^{-1}]_{ln} [\ubar{\tilde{\bf V}} \ubar{\tilde{\bf M}}^{-1} \ubar{\tilde{\bf P}}^{-1} ]_{jk} \Lambda_{i}^{T_f} \Lambda_{l}^{T_f} },
\end{equation}
which is itself affected by the connections that the zealot node makes. It is difficult to analyze the analytical equations (Eqns.\ (\ref{small Tf discrete-time energy cost generic quadratic z}) and (\ref{small Tf discrete-time turning point z*})) by inspection further, and the use of numerical experiments is needed to understand how the selection of zealot-influenced nodes affect the energy cost.

\subsection{Numerical experiments}

In the results that follow, it can be assumed that the network node states are driven toward consensus value of $c=5$, while the large $T_f$ value is assumed to be $T_f=20N$, which is sufficiently large for the network dynamics to stabilize. On the other hand, the small $T_f$ regime of the discrete-time system is relatively short, and decreases with increasing number of drivers \cite{chen2021energy}. Therefore, it suffices to analyze the small $T_f$ regime energy cost with using one driver node to control the network, setting $T_f=N+1$, which for one driver, is the smallest possible \cite{chen2021energy}.


\subsubsection{Large $T_f$ regime}

\begin{figure*} 
\includegraphics[scale=0.5]{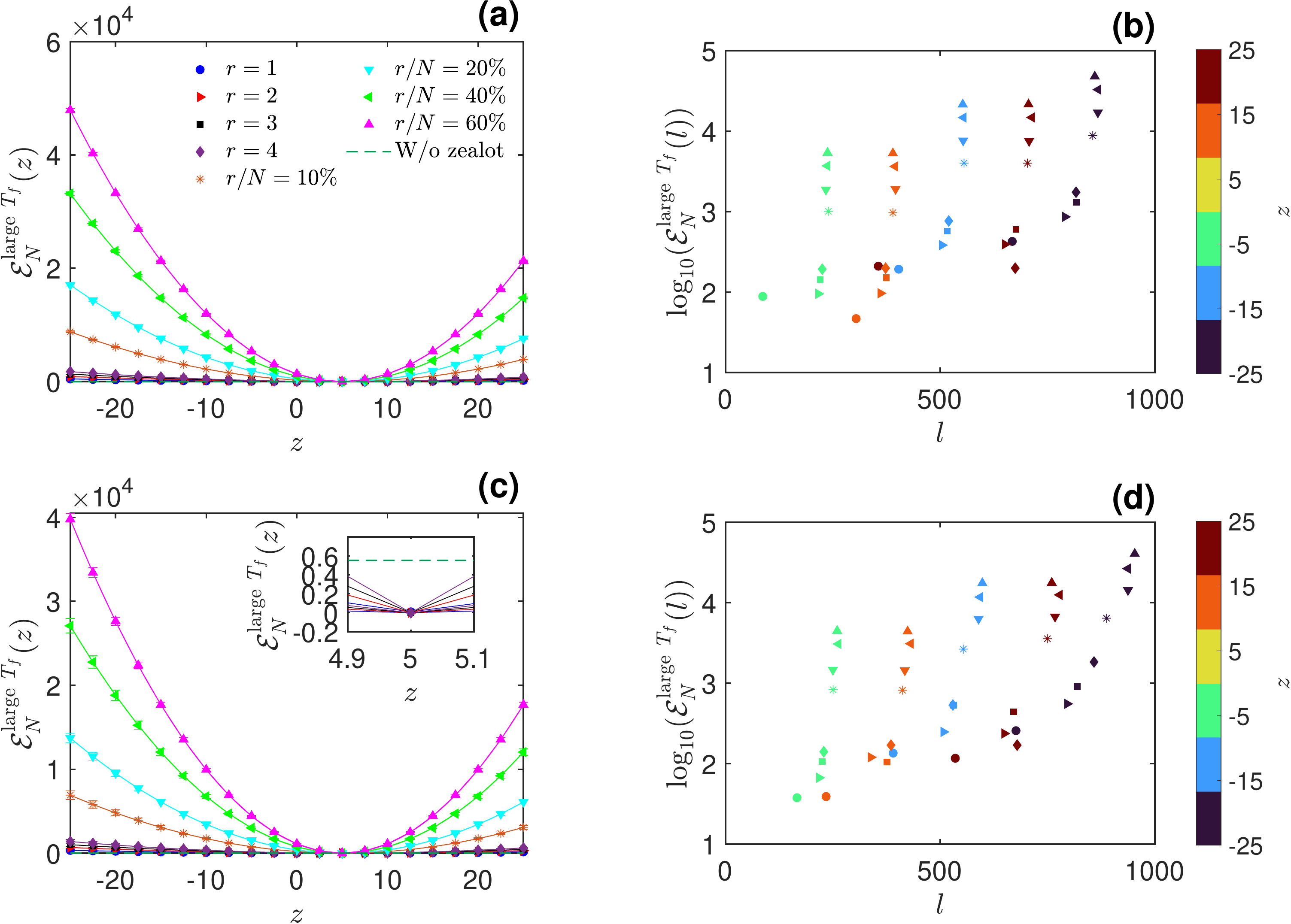} 
\caption{Conformity dynamics: Large $T_f$ regime, $N$ drivers results in $N=200$, $\langle k \rangle =6$ ER and SF networks (top and bottom panels respectively). (a) and (c) validate Eqn.\ (\ref{large Tf discrete-time energy cost generic quadratic z}), showing that the energy cost is quadratic with respect to zealot's $z$ opinion. Solid lines and markers are computed as the mean of $20$ independent selection of zealot-influenced nodes configurations, with error bars being the standard deviations. (b) and (d) show that the energy cost is proportional to the state space trajectory length $l$, with varying $r$ and $z$ values. The markers according to legend in (a) display the corresponding $r$ value, and the colors in (b) and (d) indicate the $z$ value of the influencing zealot node. Inset in (c) shows that indeed, $z^*=c=5$, with energy cost being zero.} \label{fig:fig9}
\end{figure*}

\begin{figure*}
\begin{center} 
\includegraphics[scale=0.4]{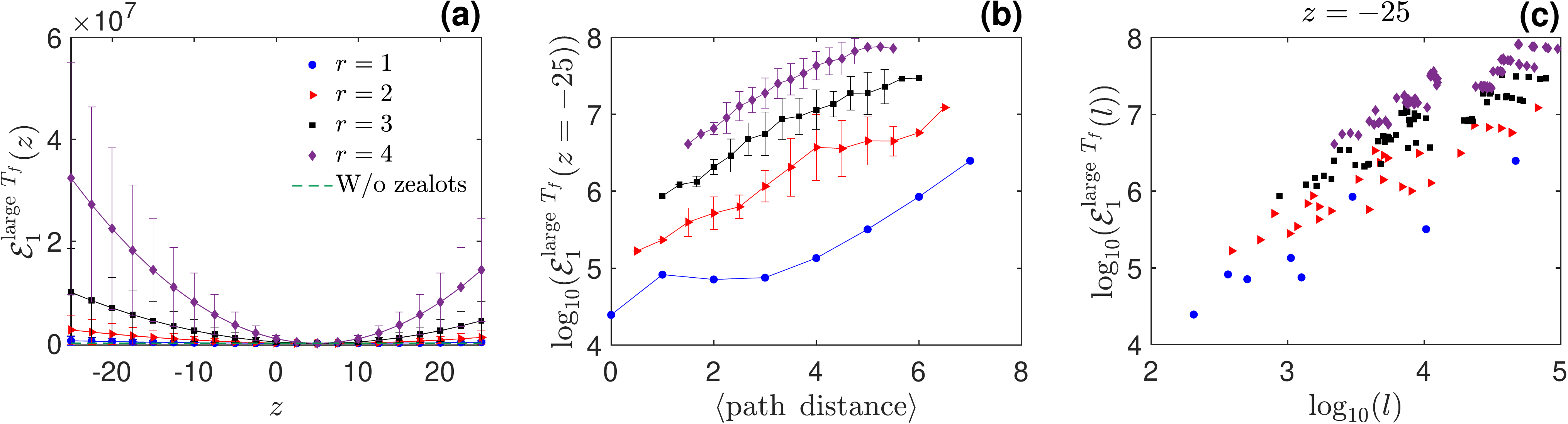}    
\end{center} 
\caption{Conformity dynamics: Large $T_f$ regime, one driver results in a $N=8$ chain network, with the driver node located at the root node (node $1$). (a) validates Eqn.\ (\ref{large Tf discrete-time energy cost generic quadratic z}), showing that the energy cost is quadratic with respect to zealot's $z$ opinion. The marker symbols denote the different $r$ number of zealot-influenced normal agents, while solid lines and markers are computed based on the mean of all possible zealot-influenced nodes configurations, with the error bars being the standard deviations. The difference in energy cost within a fixed $r$ value is attributed to where the zealot-influenced nodes are located within the network, with (b) showing that the further the nodes are away from the driver node, the more energy cost is required (assuming fixed $z=-25$). Correspondingly, (c) shows that, at fixed $z=-25$, the energy cost scales as $\log_{10}(\mathcal{E}^{\text{large }T_f}_1(l)) \sim \log_{10}(l)$. Marker symbols relate to legend in (a). } \label{fig:fig10}
\end{figure*}

\begin{figure*}
\includegraphics[scale=0.5]{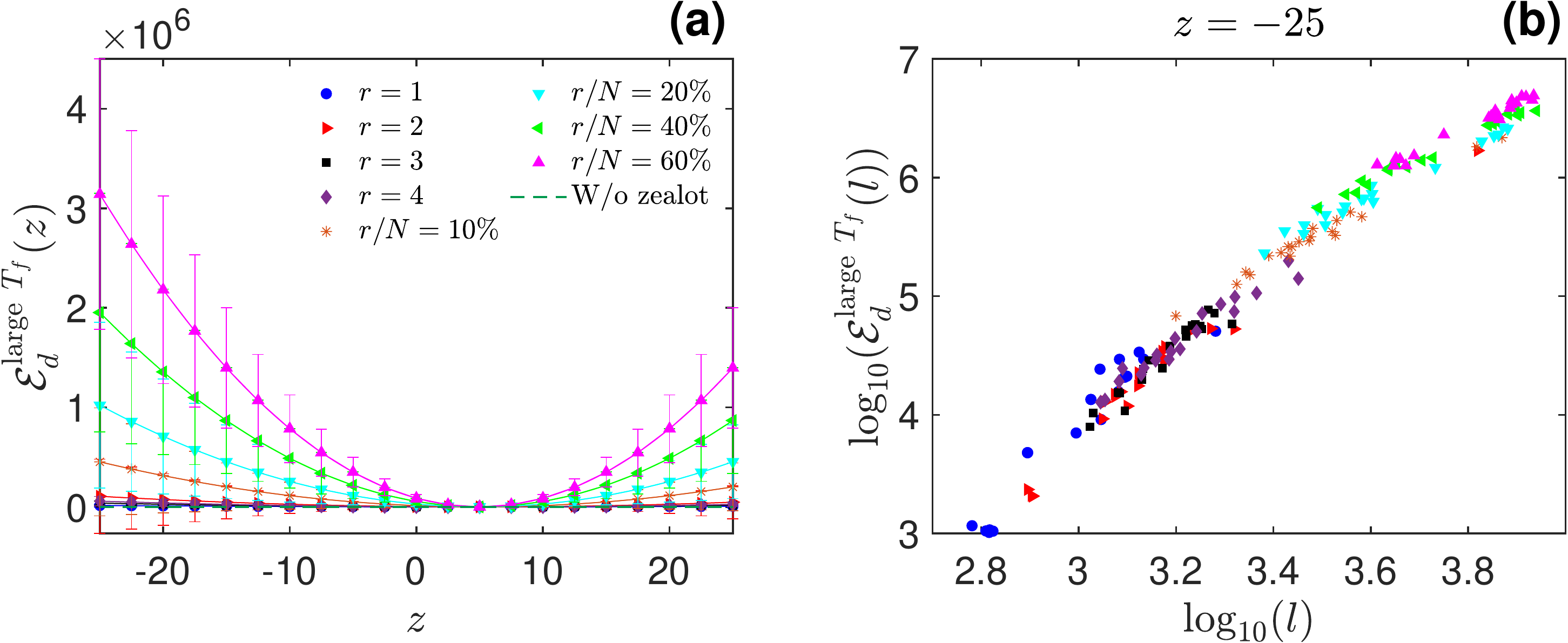} 
\caption{Conformity dynamics: Large $T_f$ regime, $d$ drivers results in a $N=200$, $\langle k \rangle = 6$ ER network. (a) validates Eqn.\ (\ref{large Tf discrete-time energy cost generic quadratic z}), showing that the energy cost is quadratic with respect to the zealot's fixed $z$ opinion. Markers and solid lines are computed based on the mean of $20$ independent selections of zealot-influenced nodes configurations, with the error bars being the standard deviations. Legend therein shows the different marker symbols corresponding to the different $r$ number of zealot-influenced nodes. (b) shows that, at fixed $z=-25$, the configurations which lead to increased state space trajectory length $l$ leads to higher energy cost required, and the energy cost scales as $\log_{10}(\mathcal{E}^{\text{large }T_f}_d(l)) \sim \log_{10}(l)$.} \label{fig:fig11}
\end{figure*}

The large $T_f$ results of the discrete-time system with conformity behavior are plotted respectively in Figs.\ \ref{fig:fig9}, \ref{fig:fig10}, and \ref{fig:fig11}, corresponding to using $N$ drivers, one driver, and $d$ drivers for control. Figs.\ \ref{fig:fig9}(a) and (c), \ref{fig:fig10}(a), and \ref{fig:fig11}(a) corroborate the predictions from Eqns.\ (\ref{large Tf discrete-time energy cost generic quadratic z}) and (\ref{large Tf conformity z turning point}) that the conformity-based system has the optimal zealot opinion at $z^*=c$, with energy cost of zero (see Fig.\ \ref{fig:fig9}(c) inset), when the zealot's opinion is the same as the consensus $c$ that the state vector is being driven toward. In the large $T_f$ limit, the zealot's fixed $z^*$ opinion steers the state vector of the network and consensus is reached, therefore costing the driver nodes no energy at all. This is regardless of the $r$ number of zealot-influenced nodes, because the conformity mechanism is global (see Fig.\ \ref{fig:fig9}(a)), and in the large time limit, one zealot-influenced node is enough to bring the entire state vector toward consensus $c$. Contrasting this with the energy cost needed to control the network in the absence of zealots (see Eqn.\ (\ref{no-zealot energy cost})), which are respectively, numerically, $1.04$, $0.554$, $36.6$, and $2.55$ for Figs.\ \ref{fig:fig9}(a) and (c), \ref{fig:fig10}(a), and \ref{fig:fig11}(a), suggesting that while the zealots have the potential to lower the energy cost to zero, its improvement over the no-zealot energy cost is marginal as compared to its continuous-time large time limit counterparts which has the potential to reduce energy cost by several orders of magnitudes. 

Further, in each of these figures, the $r$ number of zealot-influenced nodes are varied (denoted by the marker symbols therein), as well as the configurations of zealot-influenced nodes. In Figs.\ \ref{fig:fig9}(a) and (c), each $r$ value is represented by $20$ independent selections of zealot-influenced nodes chosen at random. For the $N$ drivers large $T_f$ results, the zealot-influenced nodes configurations accounted for small variations in the energy cost, denoted by the error bars plotted. Correspondingly, the relationship between the zealot-influenced state space trajectory length, as exampled in Fig.\ \ref{fig:fig9}(i), is displayed in Figs.\ \ref{fig:fig9}(b) and (d). In these figures, each $r$ value is represented by the different symbol markers according to legends in Fig.\ \ref{fig:fig9}(a), and the marker colors represent the $z$ value of the influencing zealot node. For each $r$ value, because the energy cost variations due to the different nodes configurations is little, only one random configuration was chosen. It can be seen that generally, the energy cost is proportional to length $l$:
\begin{equation}
\log_{10}(\mathcal{E}^{\text{large }t_f}_N(l)) \propto l,
\end{equation}
where an increase in $r$ leads to a slight increase in $l$ and steep increase in $\log_{10}(\mathcal{E}^{\text{large }t_f}_N(l))$; further, an increase in $z$ leads to an increase in $l$ and $\log_{10}(\mathcal{E}^{\text{large }t_f}_N(l))$, where the $z$ value that is furthest away from the turning point $z^*=c$ would cost the most energy, requiring the drivers to steer the state vector back toward consensus $c$. All in all, the results suggest that when more nodes are influenced by the zealots with strong contrarian opinions, the $N$ drivers have to consume the most amount of energy to steer the state vector back toward consensus $c$ in the large $T_f$ limit. Finally, the discrete-time length $l$ used in these computations is defined as:
\begin{equation}\label{discrete-time SS trajectory length l}
l = \sum_{\tau=0}^{T_f-1}\sqrt{\sum_{j=1}^{N}\Big[x_{j}(\tau)-x_{j}(\tau +1)\Big]^2},
\end{equation}
where index $j$ iterates over the dimensionality (or nodes) $1$ to $N$, and length $l$ is computed over the sum of the Euclidean distances of the state vector at time $\tau$ and at time $\tau+1$.


\begin{figure*}
\begin{center} 
\includegraphics[scale=0.64]{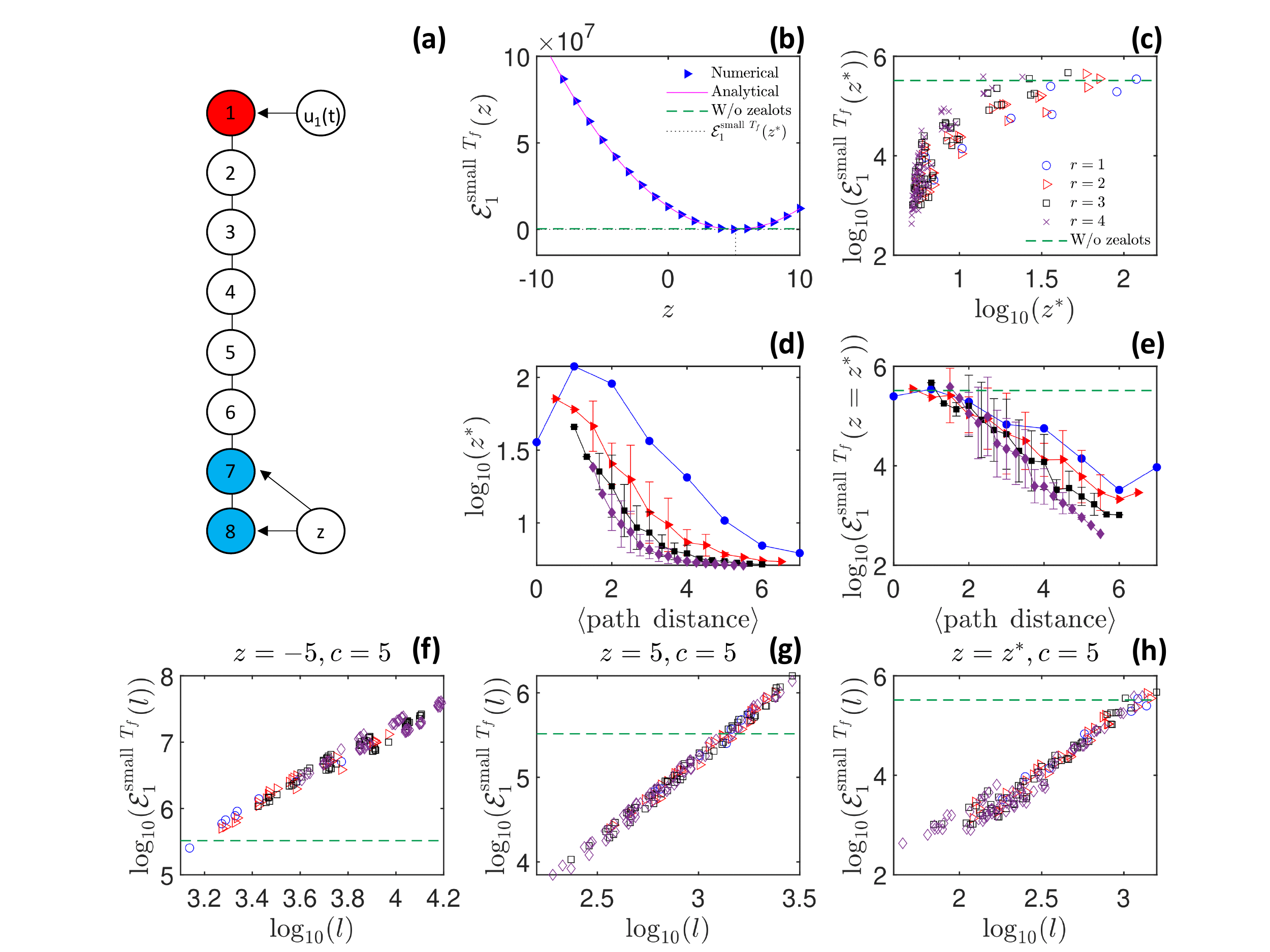} 
\end{center}
\caption{Conformity dynamics: Small $T_f$ regime, one driver results in a $N=8$ chain network, with the driver node located at the root node (node $1$). (b) validates Eqn.\ (\ref{small Tf discrete-time energy cost generic quadratic z}), showing that the energy cost is quadratic with respect to the zealot's fixed $z$ opinion. For each zealot-influenced nodes configuration, there is a different $z^*$ turning point. Shown in (c), configurations which result in low $z^*$ has associated non-zero low minima energy cost $\mathcal{E}^{\text{small }T_f}_1(z^*)$. The turning points $z^*$ and minima energy cost $\mathcal{E}^{\text{small }T_f}_1(z^*)$ are explained by where in the network are the zealot-influenced nodes located. The further away the influenced nodes are from the driver node, the lower their turning points $z^*$ and minima energy cost, as evidenced respectively in (d) and (e). (f), (g), and (h) show respectively for $z=-5$ (far away from $z^*$), $z=5$ (some configurations near $z^*$), and $z=z^*$ (all configurations at $z^*$), that the configurations which lead to increased state space trajectory length $l$ leads to increased energy cost. Marker symbols denote $r$ values following legend in (c), while green dashed line indicate the energy cost needed for control in the absence of zealots. } \label{fig:fig12}
\end{figure*}

When controlling the network with just a single control signal, it is difficult to explain the effects of the zealot in a complex network. Therefore, for this setup, the numerical experiment is performed on a simple chain network with $8$ normal agent nodes, with the driver node located at node $1$ (see cartoon representation in Fig.\ \ref{fig:fig12}(a)). In Fig.\ \ref{fig:fig10}(a), the different $r$ number of zealot-influenced nodes are denoted by the different markers according to legends therein. The error bars account for the energy cost variations due to the different configurations of zealot-influenced nodes, where an exhaustive $\Comb{N}{r}$ combinations were selected. Much like its continuous-time counterparts, the network property that explain the energy cost the best is the average path distances from diver node $1$ to the zealot-influenced normal agent nodes. At $z=-25$ (far away from the turning point $z^*=c$), it is shown in Fig.\ \ref{fig:fig10}(b) that an increase in $\langle \text{path distance} \rangle$ of the zealot-influenced nodes leads to an increase in energy cost. Further, an increase in $r$ value leads to an increase in energy cost. This suggests that the driver node has to consume the most amount of energy to steer the zealot-influenced nodes furthest away and bring it back toward consensus $c=5$. In addition, when more nodes are influenced by the zealot node, the more effort is needed to steer those nodes back toward $c=5$.

The state space trajectory length $l$ (Eqn.\ (\ref{discrete-time SS trajectory length l})) is telling of the required energy cost. At fixed zealot opinion $z=-25$, length $l$ scales with the energy cost $\mathcal{E}^{\text{large }T_f}_1(l)$ as
\begin{equation}
\log_{10}(l) \sim \log_{10}(\mathcal{E}^{\text{large }T_f}_1(l)),
\end{equation} 
where the configurations of zealot-influenced nodes which leads to an increase in $\log_{10}(l)$ leads to a linear increase in $\log_{10}(\mathcal{E}^{\text{large }T_f}_1(l))$, as shown in Fig.\ \ref{fig:fig10}(c). Further, there is a clear separation of the data points when varying the $r$, where an increase in $r$ leads to a linear increase translation in the $\log_{10}(l) \sim \log_{10}(\mathcal{E}^{\text{large }T_f}_1(l))$ plots.

The results when using $d$ drivers to control a complex network are similar. While the path distances from driver nodes to their respective nearest zealot-influenced nodes should affect the energy cost, as corroborated in Fig.\ \ref{fig:fig10}(b), this relationship is difficult to show due to many other competing terms in this setup. In Fig.\ \ref{fig:fig11}(a), each $r$ value corresponds to $20$ independent realizations of zealot-influenced nodes configurations chosen randomly, and the error bars correspond to the variation in energy cost due to the different nodes configurations. From the same configurations chosen in Fig.\ \ref{fig:fig11}(a), the state space trajectory length $l$ and associated energy cost $\mathcal{E}^{\text{large }T_f}_d(l)$ are plotted in Fig.\ \ref{fig:fig10}(b). Similar to Fig.\ \ref{fig:fig10}(c), the energy cost in Fig.\ \ref{fig:fig11}(b) scales with length $l$ as
\begin{equation}
\log_{10}(l) \sim \log_{10}(\mathcal{E}^{\text{large }T_f}_d(l)),
\end{equation} 
where a zealot-influenced nodes configuration that leads to increased state space trajectory length $l$ leads to an increase in energy cost at fixed zealot opinion $z=-25$.

\subsubsection{Small $T_f$ regime}

The results for controlling a chain network in the small $T_f$ regime using a single control signal are in Fig.\ \ref{fig:fig12}. Fig.\ \ref{fig:fig12}(b) validates Eqn.\ (\ref{small Tf discrete-time energy cost generic quadratic z}), showing that the the energy cost is quadratic with respect to the zealot's $z$ opinion. In the small $T_f$ regime, the zealot node would not have enough time to affect the entire network sufficiently, and the $z^*$ turning point is not at consensus $c$, as predicted from Eqn.\ (\ref{small Tf discrete-time turning point z*}). Further, as shown in Fig.\ \ref{fig:fig12}(c), a decrease in turning point $z^*$ leads to a decrease in minima energy cost at $z^*$, although there is no clear distinction between $r$ number of zealot-influenced nodes. This behavior is similar to its continuous-time counterparts, although it is worth pointing out that the turning points $z^*$ of the discrete-time system are all lying in the positive $z$ region. Thus, for the discrete-time dynamical system, a contrarian negative $z$ opinion (relative to positive $c$) always increases the energy cost. The zealot-influenced nodes configurations were exhaustive with all $\Comb{N}{r}$ combinations selected.

The turning points $z^*$ are related to the average path distances of the zealot-influenced nodes from driver node $1$. In Fig.\ \ref{fig:fig12}(d), an increase in $\langle \text{path distance} \rangle$ leads to a decrease in $z^*$, which correspondingly leads to decrease in associated minima energy cost $\mathcal{E}^{\text{small }T_f}_1(z=z^*)$, as shown in Fig.\ \ref{fig:fig12}(e). Thus, nodes which are far away from the driver nodes which receive influence from zealots with optimal assisting opinion $z^*$ has the most potential to reduce control energy, as compared to the no-zealot energy cost (green dashed line in Fig.\ \ref{fig:fig12}(e)), by up to a few orders of magnitude. This is in contrast to the large $T_f$ discrete-time results, where the reduction in control energy is marginal.

Correspondingly, the state space trajectory length $l$ with different fixed zealot opinions at $z=-5$, $z=5$, and $z=z^*$ are plotted respectively in Figs.\ \ref{fig:fig12}(f), (g), and (h) against the associated control energy $\mathcal{E}^{\text{small }T_f}_1(l)$. Consequently, the data points show that length $l$ scales with energy cost as
\begin{equation}
\log_{10}(l) \sim \log_{10}(\mathcal{E}^{\text{small }T_f}_1(l)),
\end{equation}
where an increase in $\log_{10}(l)$ leads to an increase in $\log_{10}(\mathcal{E}^{\text{small }T_f}_1(l))$. At $z=-5$, far away from the turning points, all configurations of zealot-influenced nodes increases the control energy relative to the no-zealot energy cost (green dashed line). At $z=5$, where some of the configurations are close to their turning points $z^*$, the zealot's presence assists in reducing the energy cost in some cases. Exactly at $z=z^*$, all configurations assists in reducing energy costs, and the data points are all below the green dashed line.

\section{Discussion} \label{discussion section}

In this paper, how zealots affect the energy cost needed to control complex social networks was examined. To do so, target control \cite{klickstein2017energy} was used, where the driver nodes are used to steer $N$ normal agents in a $n=N+1$ full network. By disallowing the zealot node to receive any directed links (from driver nodes and other normal agents), the complex dynamical system is modelled with $N$ normal interacting nodes, where there are mutual flows of information occurring, as well as a zealot node that stays fixed on their $z$ opinion trying to influence other normal agents. Thus, the energy cost is analytically derived, and it was found that for controlling social networks toward fixed $c$ final consensus state vector, the energy cost is quadratic with respect to the strongness of the zealot's $z$ opinion. In some situations where the zealots hold opinion around $z=z^*$ turning point, the zealots' presence may assist the driver nodes in reducing the control energy as compared to the situation where there were no zealots present. Away from the turning points, the zealots' presence affect the goal of controlling the network negatively and increases energy cost at a rate of $z^2$. 

Beyond the strongness of the zealot's $z$ opinion, the interplay between number of drivers, final control time regimes, network effects, network dynamics, $r$ number and configurations of zealot-influenced nodes also change the energy cost behavior. For example, when using $N$ drivers to control the network, each normal agent directly receives a control signal, and the indirect manipulations of node states through indirect paths throughout the network is decoupled from the energy cost. In this case, the zealot's influence is linear: As more nodes become influenced by the zealot, the zealot's presence has higher potential to affect the energy cost (increasing or decreasing depending on $z$) relative to the situation where there were no zealots present. In the large $t_f$ regime, node degree $\langle k \rangle$ of zealot-influenced nodes affect the turning points $z^*$ and associated minima energy cost $\mathcal{E}^{\text{large }t_f}_N(z^*)$. Nodes that have high degree $\langle k \rangle$, when influenced by the zealot helps to propagate the zealot's opinion and confer higher potential to affect the energy cost. Therefore, in SF networks where there are hubs, they are important nodes that the zealot should influence. In the small $t_f$ regime, because there is barely enough time for the zealot's opinion to take root, the choice of which nodes to influence do not matter, and only the $r$ number of nodes to be influenced is of importance.

When using less than $N$ drivers such that some nodes receive node states alteration through indirect pathways, the behavior of how zealots affect the energy cost is less obvious. For example, in some configurations, the optimal zealot opinion $z^*$ could lie in the negative $z$ region when trying to control state vector to consensus $c=5$, suggesting that a contrarian zealot may actually be beneficial for reducing energy cost. Furthermore, in most cases, the data points corresponding to different $r$ values overlap and so the energy cost is not linear with respect to $r$. Thus, a
particular $r=1$ configuration with contrarian $z^*$ may have the same potential to affect the energy as much as another $r=4$ or $r/N=20\%$ configuration. A simple explanation to understand how the energy cost is affected can be found in a simple chain network where there is only a single control signal located at the root node. For this particular setup, normal agents which are far away from the driver node are important choices for the zealot node to influence, and confer the most potential to affect the energy cost. For other types of networks, where the networked interactions are more complex, it is difficult to pinpoint any one particular network property that explains the energy cost. Nonetheless, the state space trajectory length $l$ offers a good explanation. Configurations that causes the node states to travel longer $l$ will require higher energy cost. 

How zealots affect the energy cost also change when the network dynamics change. When the node states evolve in time with conformity behavior, the most optimal zealot opinion is always $z^*=c$ in the large $T_f$ regime. This is because the conformity behavior of the discrete-time model is global, and so the zealot node can force the entire network to take its opinion. When the zealot's opinion coincides with the control goal of driving state vector toward consensus $c$, the zealot's forcing helps the driver nodes to achieve their tasks, and they do not have to take any action so no energy cost is required. Regardless of number of drivers, the zealot's influence on the energy cost is always linear with respect to $r$ number of zealot-influenced nodes in the large $T_f$ regime. For example, with any number of drivers, influencing more normal agent nodes will always confer the zealot node higher potential to affect the energy cost. Besides, the data points corresponding to different $r$ values generally do not overlap, with clear segregation of data points according to different $r$ values. In the small $T_f$ regime, however, because there is not enough time for the zealot's opinion to dominate, $z^*$ is not always $c$ and varies as the nodes configuration changes. 

Accordingly, the analytical energy cost equations with respect to parameter change is summarized in Table \ref{tab:1}.

\begin{table*}[t]
  \centering
  \caption{Summary of analytical energy cost equations as network dynamics, number of drivers, and final control time regimes change. The symbols $\ditto$ indicate that $\mathcal{E}_1^{\text{small }t_f}$, $\mathcal{E}_d^{\text{small }t_f}$, and $\mathcal{E}_d^{\text{large }t_f}$ have the same analytical equation (Eqn.\ (\ref{E_1 small t_f and E_d})). } 
\begin{tabular}{ | c | c | c | c | c | } 
 \hline
 Dynamics (number of drivers) & Small $t_f$/Small $T_f$ & Eqn.\ & Large $t_f$/Large $T_f$ & Eqn.\ \\
  \hline 
  \hline
Continuous-time ($1$ driver) & $\begin{aligned}  & c^2 \sum_{i=1}^N \sum_{j=1}^N [\tilde{\bf P}\tilde{\bf M}^{-1} \tilde{\bf P}^T]_{ij}\\
& - 2cz\sum_{i=1}^N \sum_{j=1}^N [e^{{\bf A}^T t_f}]_{ni} [\tilde{\bf P}\tilde{\bf M}^{-1} \tilde{\bf P}^T]_{ij}\\
& + z^2  \sum_{i=1}^N \sum_{j=1}^N [e^{{\bf A}^T t_f}]_{ni}[\tilde{\bf P}\tilde{\bf M}^{-1} \tilde{\bf P}^T]_{ij}[e^{{\bf A}t_f}]_{jn}\end{aligned}$ &(\ref{E_1 small t_f and E_d}) & $ \begin{aligned}
& c^2 \sum_{i,j,l,m} \tilde{p}_{li} \tilde{p}_{mj} \frac{-4\lambda_i \lambda_j}{\tilde{p}_{hi}\tilde{p}_{hj}(\lambda_i+\lambda_j)}\\
&\times \prod_{\substack{k=1\\k\neq i}}^{N}\frac{\lambda_i+\lambda_k}{\lambda_i-\lambda_k} \prod_{\substack{k=1 \\ k \neq j}}^{N}\frac{\lambda_j+\lambda_k}{\lambda_j-\lambda_k}\\
&-2cz \sum_{i,j,l,m} [{\bf V}^{-1}]_{nl} \tilde{p}_{li} \tilde{p}_{mj} \frac{-4\lambda_i \lambda_j}{\tilde{p}_{hi}\tilde{p}_{hj}(\lambda_i+\lambda_j)}\\
& \times \prod_{\substack{k=1\\k\neq i}}^{N}\frac{\lambda_i+\lambda_k}{\lambda_i-\lambda_k} \prod_{\substack{k=1 \\ k \neq j}}^{N}\frac{\lambda_j+\lambda_k}{\lambda_j-\lambda_k}\\
&+ z^2 \sum_{i,j,l,m} [{\bf V}^{-1}]_{nl}\tilde{p}_{li}\tilde{p}_{mj}p_{mn} [{\bf P}^{-1}]_{nn}\\
& \times \frac{-4\lambda_i \lambda_j}{\tilde{p}_{hi}\tilde{p}_{hj}(\lambda_i+\lambda_j)}\prod_{\substack{k=1\\k\neq i}}^{N}\frac{\lambda_i+\lambda_k}{\lambda_i-\lambda_k}\prod_{\substack{k=1 \\ k \neq j}}^{N}\frac{\lambda_j+\lambda_k}{\lambda_j-\lambda_k} \end{aligned}$ & (\ref{E_1 large tf})  \\
\hline
 Continuous-time ($d$ drivers) & \ditto & (\ref{E_1 small t_f and E_d}) & \ditto & (\ref{E_1 small t_f and E_d}) \\
 \hline
Continuous-time ($N$ drivers)   & $ c^2 t_f^{-1} N - 2czr + z^2 r t_f$ & (\ref{E_N small tf})&  $\begin{aligned} &  -2c^2 \sum_{i,j,k} \tilde{p}_{ik}\tilde{p}_{jk}\lambda_k +4cz \sum_{i,j,k} [{\bf V}^{-1}]_{ni}\tilde{p}_{ik}\tilde{p}_{jk}\lambda_k\\
&-2z^2 [{\bf P}^{-1}]_{nn} \sum_{i,j,k} [{\bf V}^{-1}]_{ni} \tilde{p}_{ik}\tilde{p}_{jk} p_{jn} \lambda_k\end{aligned}$ & (\ref{E_N large tf})\\
\hline
\hline
Discrete-time with conformity &$\sum\limits_{i=1}^{N} \sum\limits_{j=1}^{N} [\tilde{\ubar{\bf V}} \tilde{\ubar{\bf M}}^{-1} \tilde{\ubar{\bf P}}^{-1} ]_{ij} \big(c-z \big)^2 $& (\ref{large Tf discrete-time energy cost generic quadratic z}) & $\begin{aligned} & c^2 \sum_{i,j}  [\tilde{\ubar{\bf V}} \tilde{\ubar{\bf M}}^{-1} \tilde{\ubar{\bf P}}^{-1}]_{ij}\\
& - 2cz \sum_{i,j,k}\ubar{v}_{ni}[\ubar{\bf V}^{-1}]_{ij}[\tilde{\ubar{\bf V}} \tilde{\ubar{\bf M}}^{-1} \tilde{\ubar{\bf P}}^{-1}]_{jk}\Lambda_i^{T_f}\\
& + z^2 \sum_{i,j,k,l} \ubar{v}_{ni} [\ubar{\bf V}^{-1}]_{ij}\ubar{p}_{kl} [\ubar{\bf P}^{-1}]_{ln}[\tilde{\ubar{\bf V}} \tilde{\ubar{\bf M}}^{-1} \tilde{\ubar{\bf P}}^{-1}]_{jk}\\
& \times \Lambda_i^{T_f} \Lambda_l^{T_f} \end{aligned}$ & (\ref{small Tf discrete-time energy cost generic quadratic z}) \\
 \hline
\end{tabular}
  \label{tab:1}
\end{table*}

Finally, it has to be said that the introduction of zealots to networks does not fundamentally alter the controllability of the normal agents. In terms of controllability of the normal agents, the reduced $\tilde{\bf A}$ matrix and its associated Gramian captures the controllability and intrinsic eigenvalues and energy cost needed for control. The zealot node's presence acts as an external perturbation to whichever nodes that have received influence from it. For the continuous-time linear dynamical system without conformity, when less than $N$ drivers are used to steer the state vector, owing to the indirect node states manipulation, some node states may be driven higher beyond values of consensus $c$. In this case, when such node states receive a contrarian zealot influence which assists in lowering the node states, energy cost may be reduced. For the discrete-time linear dynamical system with conformity, the intrinsic network dynamics is to stabilize at a conformed state vector. Thus, when the task is to drive the state vector from zero to consensus $c$, a negative $z$ contrarian zealot forces the state vector away from $c$ and is always adversarial for network control. 

All in all, this paper explains how zealots affect the energy cost for controlling complex social networks. The results from this paper show that the interplay between the different parameters such as the zealots' beliefs, number of drivers, final control time regimes, network effects, network dynamics, number and configurations of nodes influenced by the zealots lead to different complex behavior of energy cost. Thus, when controlling a complex social network, where there may be some members who have their own biases, the energy cost behavior is altered non-trivially, suggesting that the understanding of how to control complex social systems requires a nuanced approach, and caution should be exercised when modelling real networked social systems with controlling networks with linear dynamics. Within the literature of socio-physics, there may yet exist a wealth of features that could be introduced to the controllability of complex networks that can lead to different rich phenomena. For example, Luddites are agents who actively oppose a particular idea \cite{mellor2015influence}, and the Deffuant model \cite{deffuant2000mixing, castellano2009statistical} is a network whose dynamics are linear, where agents with similar opinions communicate, leading to formation of clusters of agents with the same opinion over time, also known as echo chambers \cite{baumann2020modeling}.

\section*{Supplementary material}
Supplementary material details the derivations of the analytical equations presented.

\section*{Acknowledgements}
H.C. and E.H.Y. acknowledge support from Nanyang Technological University, Singapore,
under its Start Up Grant Scheme (04INS000175C230).

\begin{appendix}


\section{Numerical experiments for controlling different continuous-time linear dynamical networks with one driver node in the large and small $t_f$ regimes} \label{one driver extra results}

\begin{figure*}
\begin{center} 
\includegraphics[scale=0.65]{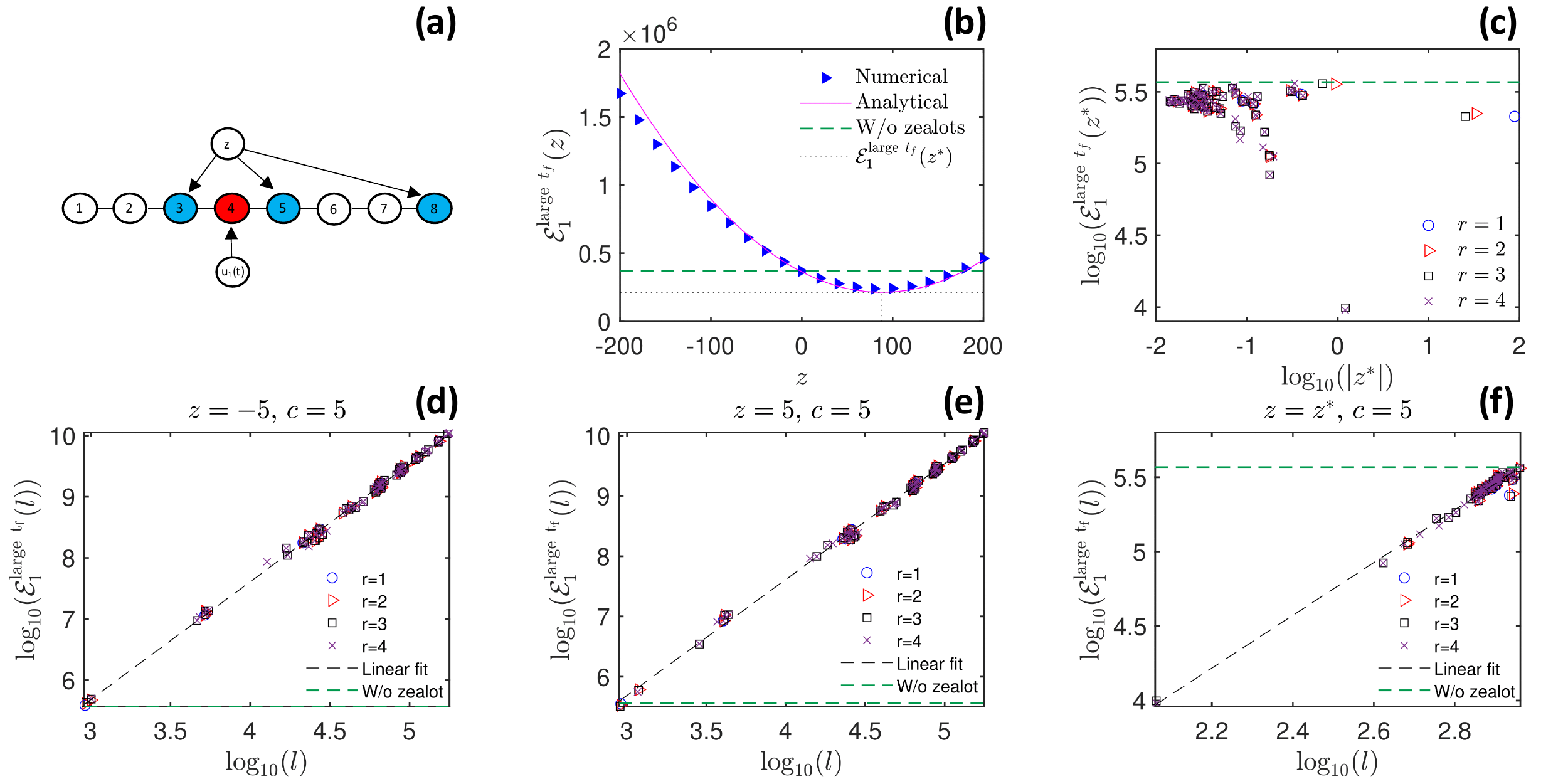} 
\end{center}
\caption{(a) Continuous-time linear dynamics: Large $t_f$ regime, one driver results in a chain network, where the driver node is located at node $4$, and the configurations of zealot-influenced nodes are varied. (b) validates Eqn.\ (\ref{E_1 large tf}). (c) shows the relationship between turning points $\log_{10}(|z^*|)$ and its associated minima energy costs $\log_{10}(\mathcal{E}^{\text{large }t_f}_1(z^*))$, and the turning points $z^*$ could lie on the negative $z$ region. Unlike Fig.\ \ref{fig:fig4}, there is no clear relationship between any network node properties such as average path distances from driver to zealot-influenced nodes, average node degree, eigenvalues, or eigencentralities to turning point $z^*$ and its minima energy costs. (d), (e), and (f) respectively indicate that the energy cost scales with associated length $l$ as $\log_{10}(\mathcal{E}(l)) \sim \log_{10}(l)$ for various configurations of zealot-influenced nodes when $z=-5$, $z=5$, and $z=z^*$ when controlling the chain network toward consensus $c=5$.  } \label{fig:fig13}
\end{figure*}

\begin{figure*}
\begin{center} 
\includegraphics[scale=0.65]{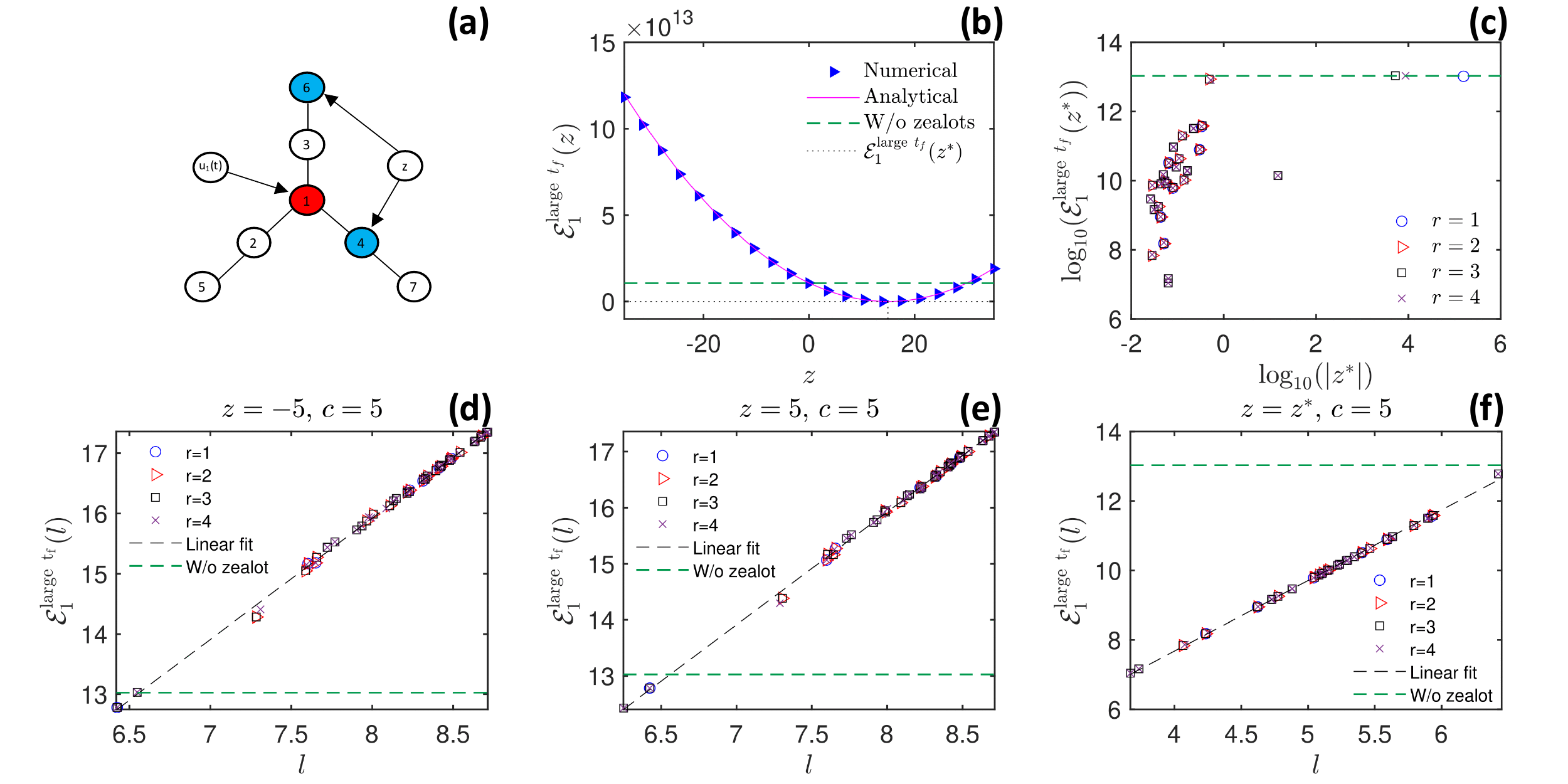} 
\end{center}
\caption{(a) Continuous-time linear dynamics: Large $t_f$ regime, one driver results in a star network, where the configurations of zealot-influenced nodes are varied. (b) validates Eqn.\ (\ref{E_1 large tf}). (c) shows the relationship between turning points $\log_{10}(|z^*|)$ and its associated minima energy costs $\log_{10}(\mathcal{E}^{\text{large }t_f}_1(z^*))$, where the turning points $z^*$ could be negative. (d), (e), and (f) respectively indicate that the energy cost scales with its associated length $l$ as $\log_{10}(\mathcal{E}(l)) \sim \log_{10}(l)$ for various configurations of zealot-influenced nodes when $z=-5$, $z=5$, and $z=z^*$ when controlling the star network toward consensus $c=5$, } \label{fig:fig14}
\end{figure*}

\begin{figure*}
\begin{center} 
\includegraphics[scale=0.65]{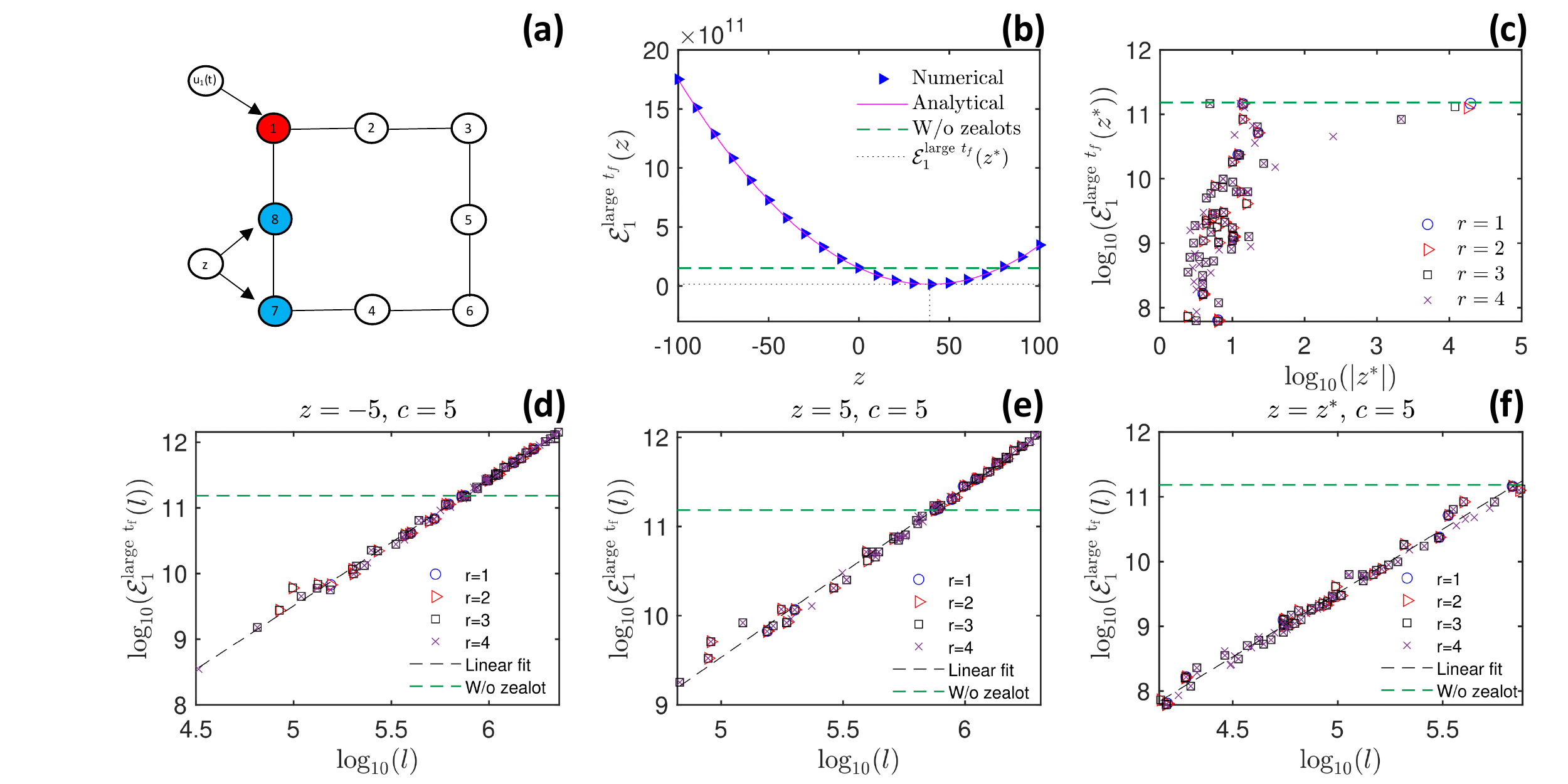} 
\end{center}
\caption{(a) Continuous-time linear dynamics: Large $t_f$ regime, one driver results in a ring network, where the configurations of zealot-influenced nodes are varied. (b) validates Eqn.\ (\ref{E_1 large tf}). (c) shows the relationship between turning points $\log_{10}(|z^*|)$ and its associated minima energy costs $\log_{10}(\mathcal{E}^{\text{large }t_f}_1(z^*))$, where the turning points $z^*$ could be negative. (d), (e), and (f) respectively indicate that the energy cost scales with its associated length $l$ as $\log_{10}(\mathcal{E}(l)) \sim \log_{10}(l)$ for various configurations of zealot-influenced nodes when $z=-5$, $z=5$, and $z=z^*$ when controlling the ring network toward consensus $c=5$. (f), (g), and (g) } \label{fig:fig15}
\end{figure*}

\begin{figure*}
\begin{center} 
\includegraphics[scale=0.65]{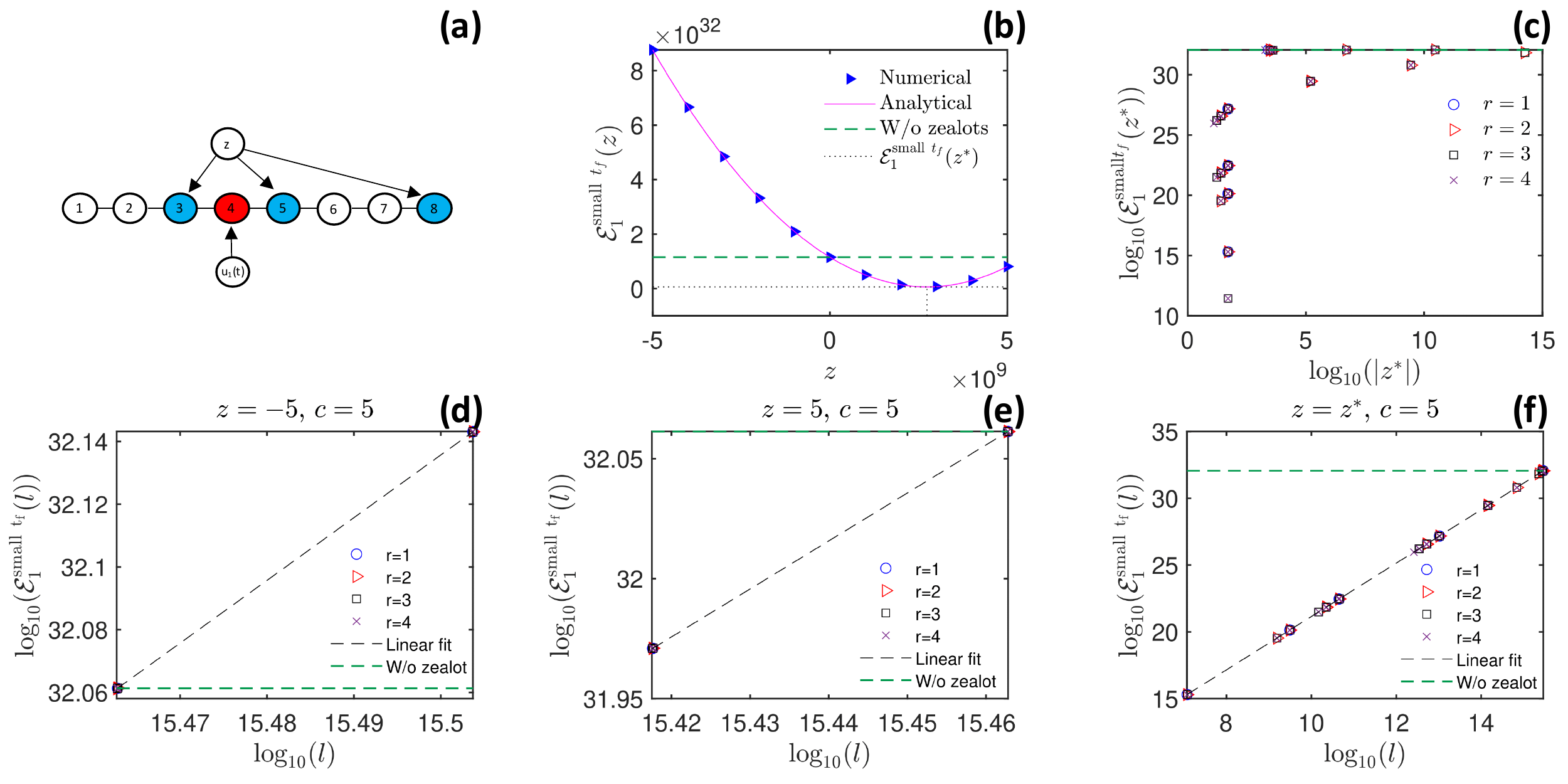} 
\end{center}
\caption{(a) Continuous-time linear dynamics: Small $t_f$ regime, one driver results in a chain network, where the driver node is located at node $4$, and the configurations of zealot-influenced nodes are varied. (b) validates Eqn.\ (\ref{E_1 small t_f and E_d}). (c) shows the relationship between turning points $\log_{10}|z^*|$ and associated minima energy costs $\log_{10}(\mathcal{E}^{\text{small }t_f}_1(z^*))$, where $z^*$ could lie in the negative region. (d), (e), and (f) respectively indicate that the energy cost scales with its associated length $l$ as $\log_{10}(\mathcal{E}(l)) \sim \log_{10}(l)$, when $z=-5$, $z=5$, and $z=z^*$ for various configurations of zealot-influenced nodes. } \label{fig:fig16}
\end{figure*}

\begin{figure*} 
\begin{center} 
\includegraphics[scale=0.65]{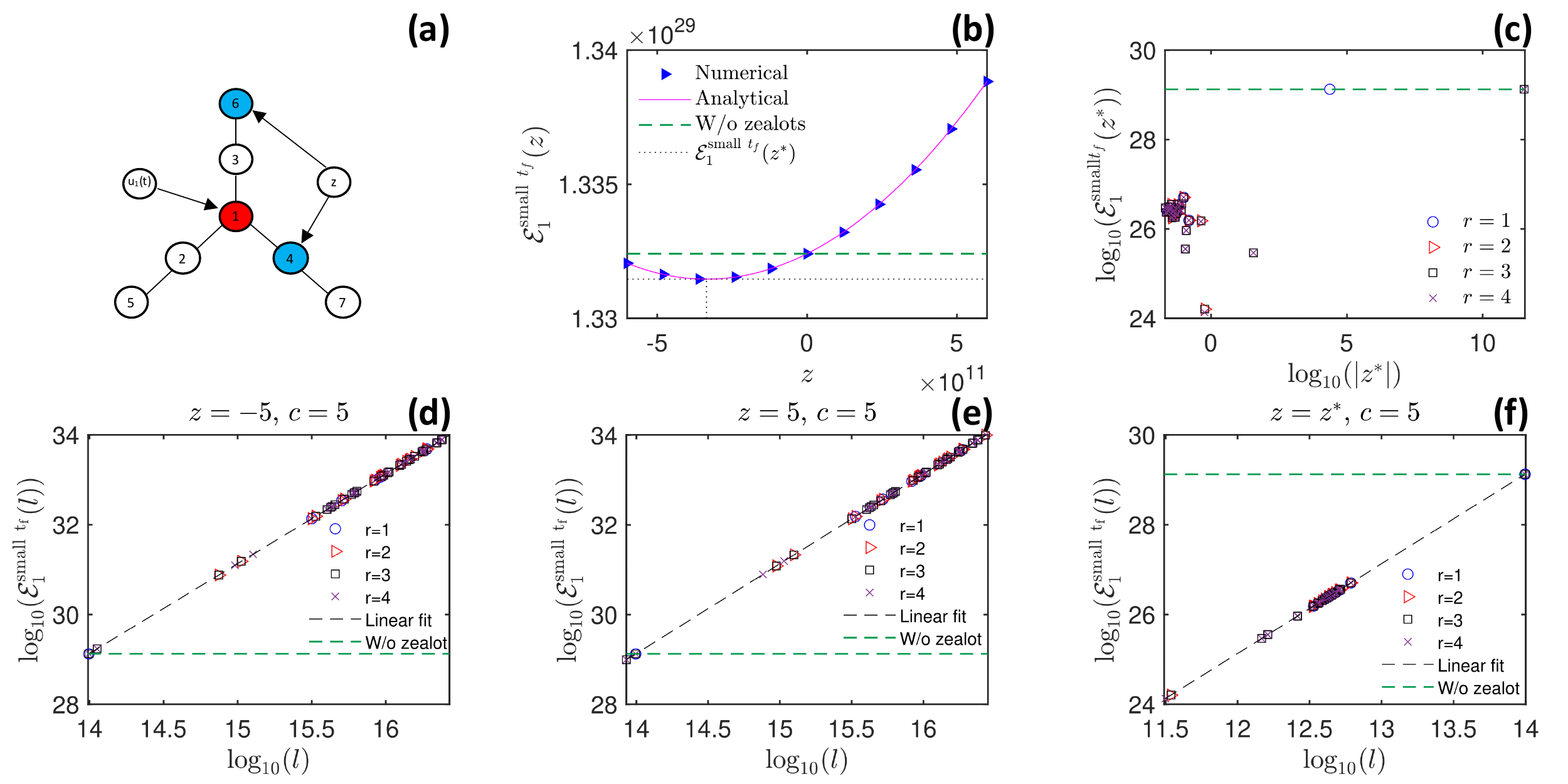} 
\end{center}
\caption{(a) Continuous-time linear dynamics: Small $t_f$ regime, one driver results in a star network, where the configurations of zealot-influenced nodes are varied. (b) validates Eqn.\ (\ref{E_1 small t_f and E_d}). (c) shows the relationship between turning points $\log_{10}|z^*|$ and associated minima energy costs $\log_{10}(\mathcal{E}^{\text{small }t_f}_1(z^*))$, where $z^*$ could lie in the negative region. (d), (e), and (f) respectively indicate that the energy cost scales with its associated length $l$ as $\log_{10}(\mathcal{E}(l)) \sim \log_{10}(l)$, when $z=-5$, $z=5$, and $z=z^*$ for various configurations of zealot-influenced nodes. } \label{fig:fig17}
\end{figure*}

\begin{figure*}
\begin{center} 
\includegraphics[scale=0.65]{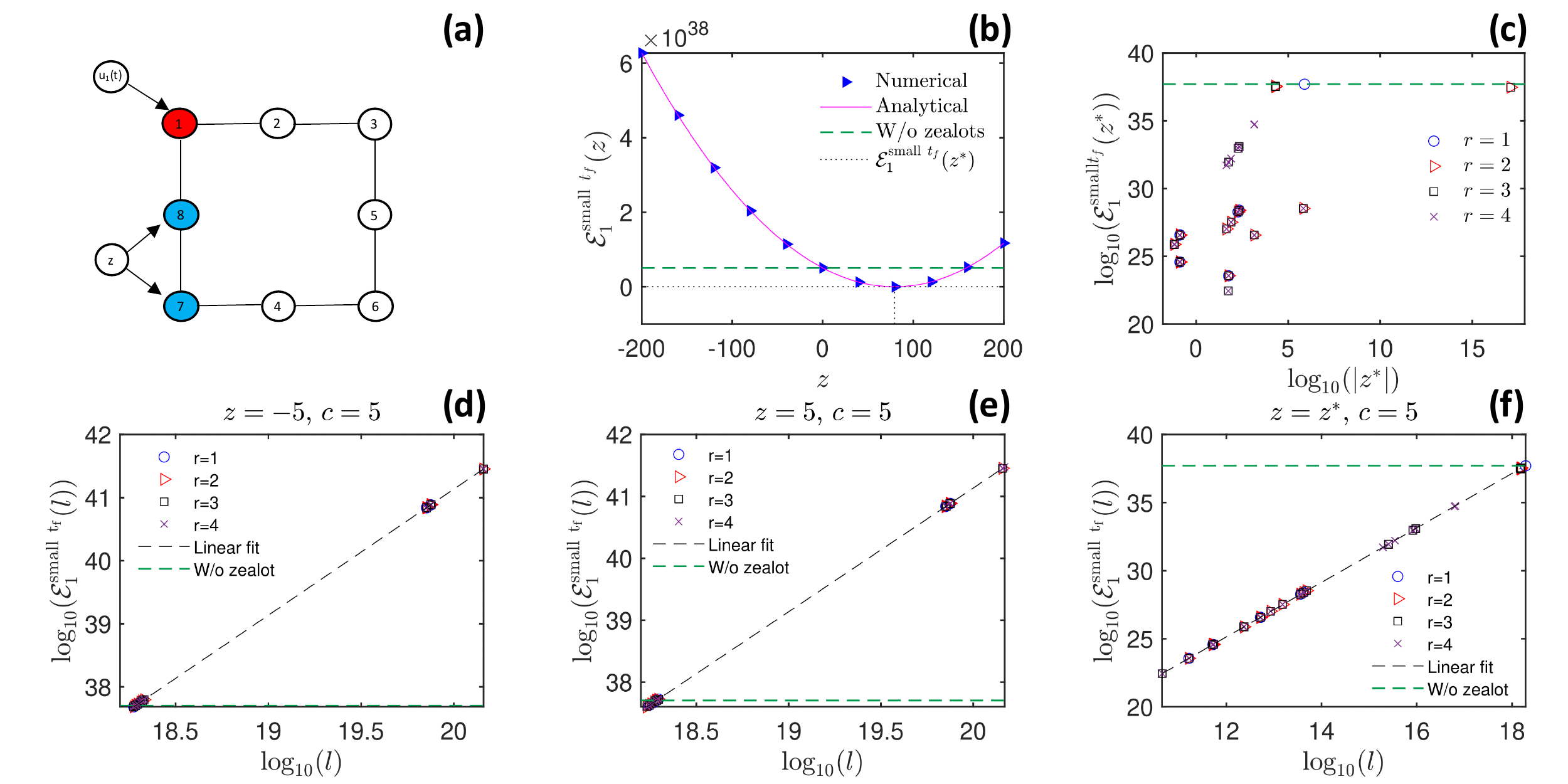} 
\end{center}
\caption{(a) Continuous-time linear dynamics: Small $t_f$ regime, one driver results in a ring network, where the configurations of zealot-influenced nodes are varied. (b) validates Eqn.\ (\ref{E_1 small t_f and E_d}). (c) shows the relationship between turning points $\log_{10}|z^*|$ and associated minima energy costs $\log_{10}(\mathcal{E}^{\text{small }t_f}_1(z^*))$, where $z^*$ could lie in the negative region. (d), (e), and (f) respectively indicate that the energy cost scales with its associated length $l$ as $\log_{10}(\mathcal{E}(l)) \sim \log_{10}(l)$, when $z=-5$, $z=5$, and $z=z^*$ for various configurations of zealot-influenced nodes.} \label{fig:fig18}
\end{figure*}

Figs. \ref{fig:fig13}\textendash\ref{fig:fig15} show the continuous-time linear dynamics numerical experiments when controlling networks with one driver node in the large $t_f$ regime respectively for chain, star, and ring networks. Figs.\ \ref{fig:fig16}\textendash\ref{fig:fig18} show likewise, except for the small $t_f$ regime.

\section{Discrete-time linear dynamics with conformity no-zealot energy cost}
In figs.\ \ref{fig:fig8}, \ref{fig:fig9}, \ref{fig:fig10}, and \ref{fig:fig11}, the green dashed lines represent the no-zealot energy cost. This computation is not the same as those in the (\ref{discrete-time LTI conformity dynamics}) dynamics and taking out the $n$-th row/column; the no-zealot network cannot be decoupled in this way owing to the ${\bf S}^{-1}{\bf A}$ coupled matrices. Instead, the no-zealot computation should consider networks in a $N\times N$ matrix, without any directed links from the $n$-th zealot node (where $n=N+1$):

\begin{equation}
\begin{aligned}
{\bf x}(\tau +1) = & \tilde{\bf S}^{-1} \tilde{\bf A} {\bf x}(\tau) + \tilde{\bf B} \tilde{\bf u}(\tau)\\
= & \ubar{\hat{\bf A}} {\bf x}(\tau) + \tilde{\bf B} \tilde{\bf u}(\tau),
\end{aligned}
\end{equation}
where ${\bf x}(\tau)=[x_1(\tau), x_2(\tau), ..., x_N(\tau) ]^T$ is a $N \times 1$ vector of the node states of $N$ normal agents, $\tilde{\bf u}(\tau)=[u_1(\tau), u_2(\tau), ..., u_M(\tau)]^T$ is the $M\times 1$ vector of input control signals, $\tilde{\bf A}$ is the network connection matrix where $\tilde{a}_{ij}$ is non-zero if there is a directed link from normal agent node $j$ to node $i$, otherwise $\tilde{a}_{ij}=0$, $\tilde{\bf B}$ describes the where in the network are the control signals attached, with $\tilde{b}_{ij}=1$ if control signal $j$ attaches to node $i$, otherwise it is zero. $\tilde{\bf S}^{-1}=\text{diag}\{\frac{1}{\tilde{s}_1}, \frac{1}{\tilde{s}_2}, ..., \frac{1}{\tilde{s}_N} \}$, where $\tilde{s}_i = \sum\limits_{j=1}^{\tilde{n}_i} \tilde{a}_{ij}$, and node $i$ has $\tilde{n}_i$ nearest neighbors. Note that $\ubar{\hat {\bf A}}$ is different from taking the $N\times N$ block from $\ubar{\bf A}$ in Eqn.\ (\ref{discrete-time LTI conformity dynamics}).

The energy cost needed to control this network is $J=\sum\limits_{\tau=0}^{T_f -1} \tilde{\bf u}^T(\tau)\tilde{\bf u}(\tau)$, and the energy-optimal control signal for controlling this system is \cite{duan2019target,lewis2012optimal}
\begin{equation} \label{no-zealot energy-optimal control signal}
\tilde{\bf u}^*(\tau)=\tilde{\bf B}^T (\ubar{\hat{\bf A}}^T)^{T_f - \tau -1} \ubar{\hat{\bf W}}^{-1}({\bf x}_f - \ubar{\hat{\bf A}}^{T_f}{\bf x}_0 ),
\end{equation}
where ${\bf x}_0=[x_1(0), x_2(0), ..., x_N(0)]^T$ is the initial state vector, ${\bf x}_f=[x_1(T_f), x_2(T_f), ..., x_N(T_f)]^T=[c,c,...,c]^T$ is the final state vector, where the node states are driven toward consensus $c$, and $\hat{\ubar{\bf W}}=\sum\limits_{\tau=0}^{T_f-1} \ubar{\hat{\bf A}}^{T_f -\tau -1}\tilde{\bf B}\tilde{\bf B}^T (\ubar{\hat{\bf A}}^T)^{T_f-\tau -1}$ is the controllability Gramian. Assuming ${\bf x}_0={\bf 0}$ and substituting Eqn.\ (\ref{no-zealot energy-optimal control signal}) into $J=\sum\limits_{\tau=0}^{T_f -1} \tilde{\bf u}^T(\tau)\tilde{\bf u}(\tau)$, the no-zealot energy cost is
\begin{equation} \label{no-zealot energy cost}
{\bf x}_f^T \hat{\ubar{\bf W}}^{-1} {\bf x}_f.
\end{equation}

In continuous-time linear dynamics, the no-zealot energy cost is the energy cost when $z=0$. However, for the discrete-time with conformity system, the no-zealot energy cost is not the same as when $z=0$. 

\section{Methods} \label{appendix methods}
For the continuous-time linear dynamics models, the $N \times N$ network connections matrix of the normal agents, $\tilde{\bf A}$, are modelled to have stable dynamics by setting self-loops for each node such that \cite{yan2015spectrum} $\tilde{a}_{ii}=-\sum\limits_{j=1}^{N}\tilde{a}_{ij}-\delta$. In this paper, $\delta=0.25$ was chosen, and all eigenvalues $\lambda_i$ are negative and the continuous-time dynamics is stable. This ensures that the eigenvalues of normal agents network are all distinct and the network is controllable with just a single control signal \cite{yan2015spectrum}.

On the other hand, the discrete-time linear dynamics with conformity models should have the diagonal entries of the first $N\times N$ block of the ${\bf A}$ matrix set to zero, such that $\ubar{\bf A}= {\bf S}^{-1} {\bf A}$ has stable dynamics with its eigenvalues $|\Lambda_i|<=1$ \cite{chen2021energy}. The minimum number of drivers needed to ensure controllability for this model is related to network nodes degree $\langle k \rangle$ \cite{wang2015controlling,chen2021energy}, where networks with sufficiently high $\langle k \rangle$ is controllable with just one driver node. This result is empirical, and high $\langle k \rangle$ is not a guarantee for controllability. Therefore, checks were done to ensure that the conformity-based network is controllable using the exact controllability method \cite{yuan2013exact} (in this research, the matlab code written for Ref.\ \cite{patel2015automated} was used). For example, if the constructed network was found to not satisfy controllability with one driver node, it was discarded and re-created again.

The model networks in this paper were constructed by the standard Erdős–Rényi random network algorithm, and the static model \cite{goh2001universal}, and the undirected link weights between normal agents, $\tilde{a}_{ij}=\tilde{a}_{ji}$, were drawn to be random uniform $(0,1]$. For the simple networks such as chain, ring, and star networks, the links are unweighted with $\tilde{a}_{ij}=\tilde{a}_{ji}=1$. In addition, checks were done to ensure that all nodes within the constructed network are reachable with at least one link (that is, there are no isolated nodes), thereby ensuring that the networks were generically controllable with a single control signal. Otherwise, the constructed network was discarded and re-created again.

The solutions to the state equations are important for calculating the node states evolution in Figs.\ \ref{fig:fig1}, \ref{fig:fig8}, and the state space trajectory lengths Eqn.\ (\ref{state space trajectory length}) and (\ref{discrete-time SS trajectory length l}). For the continuous-time linear dynamics, the solution is \cite{shirin2017optimal}
\begin{equation}
{\bf x}(t)= e^{{\bf A}(t-t_0)}{\bf x}_0 - \int_{t_0}^{t} e^{{\bf A}(t-\tau)}{\bf B}{\bf B}^T e^{{\bf A}^T(t_f-\tau)}d\tau {\bf C}^T \hat{\mathbf{\nu}}_f,
\end{equation}
where $\hat{\mathbf{\nu}}_f=-({\bf C}{\bf W}{\bf C}^T)^{-1}({\bf y}_f - {\bf C}e^{{\bf A}(t_f-t_0)}{\bf x}_0)$. This calculation was done in matlab, where the $e^{{\bf A}\tau}$ terms are especially computationally costly if the function expm(A*tau) was used. Instead, it can be sped up by eigen-decomposing the ${\bf A}$ matrix and then using P*exp(diag(D*tau))*inv(P). Furthermore, the computation $({\bf C}{\bf W}{\bf C})^{-1}$ is costly, and it is unnecessary to repeat this computation with each $t$ step. Thus, it is better to compute $({\bf C}{\bf W}{\bf C})^{-1}$ once and store as a variable to be used in each $t$ step of the numerical integration. In some calculations, for example the continuous-time one driver small $t_f$ results, the standard numerical double precision is not enough to compute the controllability Gramian and state evolution properly, thus the numerical precision has to be increased using the matlab toolbox Advanpix \cite{mct2020}. For the state equations involving the discrete-time conformity dynamics, the solution is obtained by iterating Eqn.\ (\ref{discrete-time LTI conformity dynamics}) through $\tau$.

Finally, note that for the continuous-time linear dynamics state space trajectory length $l$, the calculations were performed with time variable numerically sampled at $100$ evenly spaced values. The accuracy of the results may be improved by increasing the number of sampled numerical values. However, there is a trade-off in computation speed. Furthermore, several calculations with $1000$ evenly spaced values were computed and compared with those from $100$ evenly spaced values, and there were negligible differences.

\section{Conformity dynamics one driver analytical $\underline{\tilde{\bf M}}^{-1}$}
Assuming that the sole control signal is attached to normal agent node $h$, then adapting from Ref.\ \cite{chen2021energy}:
\begin{widetext}
\begin{equation} \label{one driver M inverse}
\ubar{\tilde{{\bf M}}}^{-1}(i,j)=\begin{cases}
\dfrac{(-1)^{(N+1)} \prod\limits_{k=1}^{N}(1-\Lambda_i \Lambda_k)\prod\limits_{\substack{l=1 \\ l\neq i}}^{N}(1-\Lambda_l \Lambda_i)}{[\ubar{\tilde{\bf P}}^{-1}]_{jh}\ubar{\tilde{v}}_{hi}\prod\limits_{\substack{k=1\\k\neq i}}^{N}(\Lambda_i - \Lambda_k) \prod\limits_{\substack{l=1\\l\neq i}}^{N}(\Lambda_l - \Lambda_i)},  & i=j\text{, }  \\
\Vast[\dfrac{(-1)^{N}}{(\Lambda_j-\Lambda_i)\ubar{\tilde{v}}_{hi}[\ubar{\tilde{\bf P}}^{-1}]_{jh}}\Vast]\
\Vast[ \dfrac{\prod\limits_{k=1}^{N}(1-\Lambda_i \Lambda_k) \prod\limits_{\substack{k=1 \\ k \neq i}}^{N}(1-\Lambda_k \Lambda_j)}{\prod\limits_{\substack{k=1 \\ k \neq i}}^{N}(\Lambda_i-\Lambda_k)\prod\limits_{\substack{k=1 \\ k \neq j,i}}^{N}(\Lambda_k-\Lambda_j)} \Vast], & i\neq j
\end{cases}
\end{equation}

\end{widetext}

\end{appendix}

\section*{References}
\bibliography{ref}

\end{document}